\documentclass[fleqn,usenatbib]{mnras}
\usepackage[utf8]{inputenc}
\usepackage{amsmath}	
\usepackage{amssymb}	
\usepackage{bm}
\usepackage{graphicx}	

\newcommand\rmi{\mathrm{i}}

\newcommand{\rhog}{{\rho_{\rm g}}}
\newcommand{\hatrhog}{{\hat{\rho}_{\rm g}}}
\newcommand{\rhogn}{{\rho_{\rm g}^0}}

\newcommand{\vg}{{{\bf v}_{\rm g}}}
\newcommand{\hatvg}{{\bf{\hat{v}}_{\rm g}}}

\newcommand{\vgx}{{v_{{\rm g}x}}}
\newcommand{\hatvgx}{{\hat{v}_{{\rm g}x}}}
\newcommand{\vgxn}{{v^0_{{\rm g}x}}}
\newcommand{\vgy}{{v_{{\rm g}y}}}

\newcommand{\vgyn}{{v^0_{{\rm g}y}}}
\newcommand{\vgz}{{v_{{\rm g}z}}}

\newcommand{\vgzn}{{v^0_{{\rm g}z}}}

\newcommand{\hatsigma}{{\hat{\sigma}}}

\newcommand{\rhod}{{\rho_{\rm d}}}
\newcommand{\rhodn}{{\rho_{\rm d}^0}}
\newcommand{\ud}{{\bf{u}}}

\newcommand{\ux}{{u_{x}}}
\newcommand{\uxn}{{u^0_{x}}}

\newcommand{\uy}{{u_{y}}}
\newcommand{\uyn}{{u^0_{y}}}

\newcommand{\uz}{{u_{z}}}
\newcommand{\uzn}{{u^0_{z}}}

\newcommand{\vd}{{\bf v}_{\rm d}}

\newcommand{\fsd}{{f_{\rm d}}}
\newcommand{\fsdn}{{f_{\rm d}^0}}
\newcommand{\fsg}{{f_{\rm g}}}
\newcommand{\fsgn}{{f_{\rm g}^0}}
\newcommand{\gsube}{{g_{\rm e}}}
\newcommand{\stokes}{{\overline{\rm St}}}
\newcommand{\rmd }{{\rm d}}  

\newcommand{\taus}{{\tau_{\rm s}}}

\newcommand{\hattaus}{{\hat{\tau}_{\rm s}}}
\newcommand{\bartaus}{{\bar{\tau}_{\rm s}}}
\newcommand{\bartausn}{{\bar{\tau}^0_{\rm s}}}

\newcommand{\wstar}{\nu^*}
\newcommand{\hatx}{{\bf \hat x}}

\defcitealias{2005ApJ...620..459Y}{YG05}
\defcitealias{2019ApJ...878L..30K}{K+19}

\title[Polydisperse Streaming Instability I]{Polydisperse Streaming Instability I. Tightly coupled particles and the terminal velocity approximation}

\author[S.-J. Paardekooper et al.]{
Sijme-Jan Paardekooper,$^{1,2}$\thanks{E-mail: s.j.paardekooper@qmul.ac.uk}
Colin P.~McNally,$^{1}$
Francesco Lovascio$^{1}$
\\
$^{1}$Astronomy Unit, School of Physics and Astronomy, Queen Mary University of London, London E1 4NS, UK\\
$^{2}$DAMTP, University of Cambridge, Wilberforce Road, Cambridge CB3 0WA, UK
}

\date{Accepted XXX. Received YYY; in original form March 2, 2020}

\begin{document}

\maketitle

\begin{abstract}
We introduce a polydisperse version of the streaming instability, where the dust component is treated as a continuum of sizes. We show that its behaviour is remarkably different from the monodisperse streaming instability. We focus on tightly coupled particles in the terminal velocity approximation and show that  unstable modes that grow exponentially on a dynamical time scale exist. However, for dust to gas ratios much smaller than unity they are confined to radial wave numbers that are a factor $\sim 1/\stokes$ larger than where the monodisperse streaming instability growth rates peak. Here $\stokes\ll 1$ is a suitable average Stokes number for the dust size distribution. For dust to gas ratios larger than unity, polydisperse modes that grow on a dynamical time scale are found as well, similar as for the monodisperse streaming instability and at similarly large wave numbers. At smaller wave numbers, where the classical monodisperse streaming instability shows secular growth, no growing polydisperse modes are found under the terminal velocity approximation. Outside the region of validity for the terminal velocity approximation, we have found unstable epicyclic modes that grow on $\sim 10^4$ dynamical time scales.
\end{abstract}


\section{Introduction}
The focus of this paper is the Streaming Instability \citep[SI,][hereafter YG05]{2005ApJ...620..459Y}, 
which we generalize from the original case of single-size dust to polydisperse dust with a continuum of particle sizes. 
The SI is a promising mechanism for building km-sized planetesimals out of cm-sized pebbles \citep[e.g.][]{2007Natur.448.1022J, 2010ApJ...722.1437B, 2016ApJ...822...55S,2018A&A...618A..75S}. Recently, \citet{2019ApJ...878L..30K}, hereafter K+19, presented results for a discrete polydisperse dust distribution with a large number of particle sizes, finding in most cases slower growth compared to the monodisperse case. For many parameters, they found very slow, but descending growth rates for instability as the number of discrete dust sizes considered increased, approaching the  continuum limit. In this paper, we approach the problem from a continuum perspective, a scenario we will refer to as PSI (Polydispserse Streaming Instability). We focus on tightly coupled particles, a case that lends itself to analytic understanding in addition to numerical calculations. In particular, the well-known terminal velocity (TV) approximation \citep[e.g.][]{2014MNRAS.440.2136L,2017ApJ...849..129L} applies to the PSI, a scenario we refer to as TV-PSI. We identify regions of parameter space where the TV-PSI shows exponentially growing modes, and compare these to the classic monodisperse SI.

The plan of this paper is as follows. In section \ref{sec:physics} we derive the equations governing a gas coupled to a solid component with a continuous size distribution. From these we obtain the linearized equations for the PSI in section \ref{sec:linear}. In section \ref{sec:TV} we detail the different wave number regimes relevant for the TV-PSI. Numerical results are presented in section \ref{sec:numerical}, and we conclude in section \ref{sec:disc}.

\section{Physical Model}
\label{sec:physics}

\subsection{Governing equations}
We are interested in the evolution of a mixture of solid particles (dust) and gas, where the two phases are coupled through a drag force. Consider the distribution function for dust particles $f({\bf x}, {\bf v}, a, t)$ so that 
\begin{align}
f({\bf x}, {\bf v}, a, t)\rmd{\bf x} \rmd{\bf v} \rmd a\, ,
\end{align}
is the number of dust particles in a volume $d{\bf x}$ around ${\bf x}$, with velocities in a (velocity) volume $d{\bf v}$ around ${\bf v}$ and with size between $a$ and $a+da$. The evolution of $f$ is given by the nonlinear Boltzmann equation
\begin{align}
\partial_t f + {\bf v}\cdot \nabla_x f + \nabla_v\cdot (f {\bf F})= 0\, ,
\label{eqBoltzmann}
\end{align}
where ${\bf F}$ is the force per unit mass acting on the dust. Of particular interest is the drag force, which we take to be in the Epstein regime: 
\begin{align}
{\bf F}_{\mathrm{drag},\mathrm{d}} = -\frac{{\bf v}-\vg}{\taus(a)}\, ,
\end{align}
where $\vg$ is the gas velocity and $\taus$ is the \emph{particle} stopping time, which is proportional to $a/\rhog$, where $\rhog$ is the gas density. Equation (\ref{eqBoltzmann}) is a simplified version of the spray equation \citep{1958PhFl....1..541W} commonly used for dilute polydisperse particle-gas flows. We do not consider changes in size, heat transfer, nucleation or collisions between particles.  

For our investigation of the PSI we take our domain to be an unstratified shearing box. By taking velocity moments of (\ref{eqBoltzmann}) one obtains fluid equations for the dust component:
\begin{align}
\partial_t\sigma + \nabla\cdot(\sigma {\bf u} )=& 0\, ,\label{eqDustCont}\\
\partial_t{\bf u}
+ ({\bf u}\cdot\nabla) {\bf u}
=&
-2\bm{\Omega}\times {\bf u}-\nabla\Phi - \frac{{\bf u}-\vg}{\taus(a)}\, ,\label{eqDustMom}
\end{align}
where $\bm{\Omega}$ is the angular velocity of the box and $\Phi = -S\Omega x^2$ is the effective potential, with $S$ the shear rate of the disc ($S=3\Omega/2$ in a Keplerian disc). The size-density $\sigma$ and velocity ${\bf u}$ are size-dependent and defined in such a way that 
\begin{align}
\rhod = \int \sigma \rmd a & \qquad
\rhod \vd = \int \sigma {\bf u} \rmd a  \, , 
\end{align}
where $\rhod$ is the dust volume density and $\vd$ is the bulk velocity of the dust component. The total amount of momentum transfer between gas and dust is simply
\begin{align}
\rhod {\bf F}_{\mathrm{drag},{\rm d}} = -\int \sigma \frac{{\bf u}- \vg}{\taus(a)}\rmd a = -\rhog {\bf F}_{\mathrm{drag},{\rm g}}\, ,
\label{eq:Fdrag}
\end{align}
where the last equality follows from momentum conservation. Note that the fluid approximation is only valid for particles for which the coupling to the gas is strong enough \citep{2004ApJ...603..292G, 2011MNRAS.415.3591J}. For a fluid treatment of polydisperse dust, we require that this be true for every particle size present.

The gas component obeys the usual shearing box equations, but with a drag force that is an integral due to (\ref{eq:Fdrag}):
\begin{align}
\partial_t\rhog + \nabla\cdot(\rhog \vg) =& 0\, ,\label{eqGasCont}\\
\partial_t \vg
+ (\vg \cdot \nabla)\vg
=&
2\eta {\bf \hat x} -\frac{\nabla p}{\rhog}- 2\bm{\Omega}\times \vg-\nabla\Phi + {\bf F}_{\mathrm{drag},{\rm g}} 
\label{eqGasMom}\, .
\end{align}
We take the equation of state for the gas to be isothermal, $p=c^2\rhog$, with sound speed $c$, and we have included a parameter $\eta$ governing the sub-Keplerian nature of the disc. While in the local model, $\eta$ is an input parameter effectively setting the length scale of the streaming instability \citepalias{2005ApJ...620..459Y}, in a global context it is related to the radial pressure gradient in the disc, $2\eta\rhog= -\partial P/\partial r$, where $P$ is the (global) pressure\footnote{Note that our $\eta$ is dimensional, and related to the definition of \citetalias{2005ApJ...620..459Y} by $\eta = r\Omega^2\eta_{\rm YG}$, where $r$ is the fiducial orbital radius of the shearing box. This choice is largely cosmetic: \citetalias{2005ApJ...620..459Y} in the end non-dimensionalize the problem using a length scale $\eta_{\rm YG} r$, while we use a length scale $\eta/\Omega^2$. The latter avoids using $r$, which is more natural in a purely local context.}. The equations governing the dynamics of the mixture are then (\ref{eqDustCont}), (\ref{eqDustMom}), (\ref{eqGasCont}) and (\ref{eqGasMom}). Discrete versions of these equations have been used in a protoplanetary disc context by  \cite{2018MNRAS.479.4187D} and \cite{2019ApJS..241...25B}.

\subsection{Terminal velocity approximation}
In the case of a monodisperse dust fluid, for tightly coupled particles a simplification is possible by assuming that all particles reach their terminal velocity \citepalias{2005ApJ...620..459Y}. One can then treat the mixture as a single fluid moving with the centre-of-mass velocity \citep{2014MNRAS.440.2136L}. For a polydisperse dust fluid, the terminal velocity (TV) equations read (see Appendix \ref{app:tv_derive} for a detailed derivation):
\begin{align}
\partial_t\rho + \nabla\cdot\left(\rho{\bf v}\right)&= 0\, ,\\
\partial_t{\bf v} + ({\bf v}\cdot\nabla){\bf v}  &= \frac{2p\eta{\bf \hat x}}{c^2\rho} - \frac{\nabla p}{\rho} -2\bm{\Omega}\times {\bf v}-\nabla\Phi\, ,\\
\partial_tp + \nabla\cdot\left(p{\bf v}\right)&= \mathcal{C}_{\rm g}\, ,\\
\partial_t\sigma + \nabla\cdot\left(\sigma {\bf v}\right)&= \mathcal{C}_{\rm d}\, ,
\end{align}
where $\rho = \rhog + \rhod$ is the total density of the mixture, ${\bf v} = (\rhog\vg + \rhod \vd)/\rho$ is the centre-of-mass velocity, and we have ``cooling terms'' \citep{2017ApJ...849..129L}:
\begin{align}
\mathcal{C}_{\rm g} &= \nabla\cdot\left(\left(1-\frac{p}{c^2\rho}\right)\frac{p}{\rho}\bartaus\left(\nabla p - \frac{2p \eta{\bf \hat x}}{c^2}\right) \right),\\
\mathcal{C}_{\rm d} &= \nabla\cdot\left(\frac{\sigma}{\rho}\left(\fsd\bartaus-\taus(a)\right)\left(\nabla p - \frac{2p\eta{\bf \hat x}}{c^2}\right)\right).
\end{align}
Here $\bartaus = \rhod^{-1}\int\sigma \taus \rmd a$ is an average stopping time, and $\fsd=\rhod/\rho$ denotes the dust mass fraction. Note that compared to the thermodynamic one-fluid TV approximation \citep{2017ApJ...849..129L}, we need an extra equation to track the evolution of the size-density $\sigma$. 

\section{Linear analysis of the PSI}
\label{sec:linear}

\subsection{Equilibrium state}

We take the background gas and dust (size-) density to be constant in space, and all velocities to be independent of $y$ and $z$ and $\vgz=\uz=0$.  Time-independent solutions can then be found where only the $y$ component of the velocities vary with $x$ according to $d_xu_y, d_x \vgy \propto -S$. The four momentum equations read, under these assumptions:
\begin{align}
\vgy=& -S x - \frac{\eta}{\Omega}   - \frac{1}{2\Omega \rhog}\int \sigma \frac{u_x- \vgx}{\taus(a)}\rmd a\, ,
\label{eq:eqgasmomx}\\
\vgx  =& \frac{1}{(2\Omega-S)\rhog}\int \sigma \frac{u_y- \vgy}{\taus(a)}\rmd a\, ,
\label{eq:eqgasmomy}\\
u_y=& -S x + \frac{u_x-\vgx}{2\Omega\taus(a)}\, ,
\label{eq:eqdustmomx}\\
u_x  =&  - \frac{u_y-\vgy}{(2\Omega-S) \taus(a)}\, ,
\label{eq:eqdustmomy}
\end{align}
Combine the two dust momentum equations to obtain an expression for the relative velocity $u_y-\vgy$:
\begin{align}
u_y - \vgy=& -\frac{\kappa^2\taus(a)^2(\vgy+S x) +(2\Omega-S)\vgx\taus(a)}{1+\kappa^2\taus(a)^2}\, ,
\label{eq:rely}
\end{align}
with epicyclic frequency $\kappa^2=2\Omega(2\Omega-S)$. Using this in the gas $y$ momentum equation (\ref{eq:eqgasmomy}) yields
\begin{align}
\vgx  =& -\frac{2\Omega\mathcal{J}_1(\vgy +S x)}{1+\mathcal{J}_0}\, ,
\label{eq:eqvgx}
\end{align}
where we have defined the integrals
\begin{align*}
\mathcal{J}_\alpha = \frac{1}{\rhog}\int \frac{\sigma (\kappa\taus(a))^{\alpha}}{1+\kappa^2\taus(a)^2} \rmd a \, .
\end{align*}
Using (\ref{eq:eqdustmomx}) and (\ref{eq:rely}) in the gas $x$ momentum equation (\ref{eq:eqgasmomx}) yields
\begin{align}
\vgy=& -S x - \frac{\eta}{\Omega}   -(\vgy+S x)\mathcal{J}_0 +\frac{\vgx\mathcal{J}_1}{2\Omega} \, ,
\end{align}
and finally using (\ref{eq:eqvgx}) we get an explicit expression for $\vgy$:
\begin{align}
\vgy
&=
-Sx -\frac{\eta}{\Omega}\frac{1 +  \mathcal{J}_0}{\left(1+ \mathcal{J}_0\right)^2 + \mathcal{J}_1^2}\, ,
\end{align}
from which the remaining velocities follow in a straightforward way from (\ref{eq:eqvgx}), (\ref{eq:rely}) and (\ref{eq:eqdustmomy}). As a result, we obtain a local shearing box analog of the equations derived by \cite{2005ApJ...625..414T} and used in \cite{2018MNRAS.479.4187D}, but generalized to arbitrary $\kappa$:
\begin{align}
\vgx
&=
\frac{2\eta}{\kappa}\frac{\mathcal{J}_1}{\left(1+ \mathcal{J}_0\right)^2 + \mathcal{J}_1^2},\label{eq:veqfirst}
\\
\vgy
&=
-Sx -\frac{\eta}{\Omega}\frac{1 +  \mathcal{J}_0}{\left(1+ \mathcal{J}_0\right)^2 + \mathcal{J}_1^2}\, ,\\
\ux &= \frac{2\eta}{\kappa}
 \frac{\mathcal{J}_1 - \kappa\taus(a)(1 +  \mathcal{J}_0)}{(1+\kappa^2\taus(a)^2)(\left(1+ \mathcal{J}_0\right)^2 + \mathcal{J}_1^2)} \,  ,\\
\uy   &= -Sx
- \frac{\eta}{\Omega}\frac{1 +  \mathcal{J}_0 + \kappa\taus(a)\mathcal{J}_1}{(1+\kappa^2\taus(a)^2)(\left(1+ \mathcal{J}_0\right)^2 + \mathcal{J}_1^2)}\,.\label{eq:veqlast}
\end{align}
In the limit of a single size dust fluid and Keplerian rotation ($\kappa=\Omega$) we recover the solution of \cite{1986Icar...67..375N}. In the TV approximation, under the same assumptions as above, the equilibrium centre-of-mass velocity is ${\bf v} = (-Sx-\fsg\eta/\Omega){\bf \hat y}$, where $\fsg=\rhog/\rho$ is the gas fraction. In all our numerical results, we take the background disc to be Keplerian with $S=3\Omega/2$ and therefore $\kappa=\Omega$.

\subsection{Linear perturbations}
Consider small perturbations such that $\rhog = \rhogn + \hatrhog\exp(\rmi {\bf k}\cdot {\bf x} - \rmi \omega t)$, where $\rhogn$ is the background state with $|\hatrhog| \ll \rhogn$, 
and similarly for other quantities, yielding:
\begin{align}
k_x \vgxn \frac{\hatrhog}{\rhogn}
+ {\bf k}\cdot \hatvg &=
\omega \frac{\hatrhog}{\rhogn} \, ,
\label{eqFullFirst}\\
k_x \vgxn \hatvg
+ \rmi S \hatvgx {\bf \hat y}
+{\bf k} c^2 \frac{\hatrhog}{\rhogn}
- 2\rmi \bm{\Omega}\times \hatvg & &\nonumber\\
+ \frac{\rmi}{\rhogn}\int \hat \sigma\frac{\Delta{\bf u}^0}{\taus(a)} \rmd a
+ \frac{\rmi}{\rhogn}\int \sigma^0\frac{{\bf \hat u} -\hatvg}{\taus(a)} \rmd a
&=
\omega \hatvg\, , \label{eqFullVg}\\
 k_xu_x^0\hat \sigma
+ \sigma^0{\bf k}\cdot {\bf \hat u} &=
\omega \hat\sigma\, , \label{eqFullSigma}\\
k_x u_x^0 {\bf \hat u}
 + \rmi S\hat u_{x}{\bf \hat y}
-2\rmi\bm{\Omega}\times {\bf \hat u} & \nonumber\\
-\rmi \frac{{\bf \hat u}- \hatvg}{\taus(a)}
-\rmi \frac{\hat\rhog}{\rhogn} \frac{\Delta {\bf u}^0}{\taus(a)}
&=
\omega {\bf \hat u}.
\label{eqFullLast}
\end{align}
These equations form an integral equation eigenvalue problem for the eigenvalue $\omega$.

\subsection{Incompressible terminal velocity approximation}
Taking the same form for the perturbations in the TV approximation, while at the same time assuming the gas to be incompressible (see Appendix \ref{app:tv_disperse} for details), we find:
\begin{align}
-\rmi \omega \hat\rho
+ \rmi {\bf k}\cdot {\bf \hat v} =& 0\, ,\label{eqPertRho}\\
-\rmi\omega{\bf \hat v}
- S\hat v_x{\bf \hat y}
=&
-\gsube{\bf \hat x}\hat\rho
- \rmi c^2 \fsgn {\bf k}\hat p -2\bm{\Omega}\times {\bf \hat v}\, ,\label{eqPertVel}\\
\rmi{\bf k}\cdot{\bf \hat v} =&
-\bartausn
\left(c^2k^2 \fsdn \fsgn \hat p
+ \rmi k_x \gsube(\fsgn -\fsdn )\hat\rho\right)\nonumber\\
&-\rmi k_x \gsube \fsdn \hattaus\, ,\label{eqPertPres}\\
\rmi\omega\hat\sigma - \rmi{\bf k}\cdot{\bf \hat v} =&
\left(\fsdn\bartausn-\taus(a)\right)\left(k^2 c^2 \fsgn\hat p + \rmi k_x\gsube \left[\hat\sigma - \hat\rho\right]\right) \nonumber\\
&+
\rmi k_x \gsube \left(\fsd \hattaus + \fsgn \bartausn \hat\rho\right),\label{eqPertSigma}
\end{align}
with $\gsube = 2\fsgn \eta$ and perturbed stopping time
\begin{align}
\hattaus &= \frac{1}{\rhodn}\int \hat\sigma \sigma^0(a)\taus(a) {\rm d}a -  \frac{\bartausn\hat\rho }{f_d^0}\, .
\label{eqTauHat}
\end{align}
Note that in a monodisperse dust fluid $\hattaus=0$. These are the equations for the linear, incompressible polydisperse streaming instability in the terminal velocity approximation, which for reasons of brevity we will refer to as TV-PSI. We note that compressibility effects for the SI are known to be small \citep{2007ApJ...662..613Y}.

\section{Terminal velocity modes}
\label{sec:TV}

In this section, we focus exclusively on the TV-PSI, and make comparisons to its monodisperse counterpart, the SI in the TV approximation.

\subsection{Dispersion relation}

Equations (\ref{eqPertRho})--(\ref{eqPertSigma}) can be combined to give a dispersion relation (for details see Appendix \ref{app:tv_disperse}): 
\begin{align}
& \left(\frac{k^2}{k_z^2}\omega^2 -\kappa^2\right) (\omega  - \fsdn\bartausn k_x \gsube)
=  \nonumber\\
&\qquad  \bartausn \fsdn 
\left(\rmi \frac{k^2}{k_z^2} (\omega^2-\kappa^2)\omega^2
- k_x \gsube \kappa^2\right)  \left(1-\mathcal{I}(\wstar )\right),
\end{align}
with $\wstar  = (\omega -\fsdn \bartausn k_x\gsube)/(k_x \gsube \bartausn)$ and integral
\begin{align}
\mathcal{I}(\wstar ) = \frac{1}{\rhodn}\int
\frac{\sigma^0(a)}{\wstar  + \frac{\taus(a)}{\bartausn}}\left(\frac{\taus(a)}{\bartausn}\right)^2 {\rm d}a.
\end{align}
Note that $\wstar =-\taus(a)/\bartausn$ signals a resonance, where the mode phase speed matches the (size-dependent) dust advection speed:
\begin{align}
\frac{\omega}{k_x} = 2f_g\eta(f_d\tau  - \tau(a))= u_x(a)\, .
\end{align}
Note that the second equality sign assumes the TV limit of the background velocity.  This resonance turns out to be important when interpreting the results. We note that this resonance is related but different from another resonance arising in the theory of classical SI, where the dust advection speed matches the propagation speed of a wave in the gas, which gives rise to the theory of Resonant Drag Instabilities \citep[RDIs,][]{2018MNRAS.477.5011S, 2018ApJ...856L..15S}. 

The term $\propto \omega^4$ on the right-hand side of the dispersion relation leads to spurious modes and therefore should be dropped \citep{2017ApJ...849..129L}. With the expression for $\mathcal{I}$, it is straightforward to show that $\wstar =0$, or $\omega=\fsdn \bartausn k_x \gsube$ is always a solution. This is a neutral mode for which the perturbation in average stopping time exactly cancels the pressure perturbation, yielding $\mathcal{\hat C}_{\rm g}=0$. Dividing out this solution, and taking the limit of a monodisperse dust fluid, for which $\mathcal{I}=1/(1+\wstar )$, we obtain the usual cubic dispersion relation of the SI (\citetalias{2005ApJ...620..459Y}, \citealt{2011MNRAS.415.3591J}). Multiplying both sides by $\wstar +1$, we find:
\begin{align}
\left(\nu^2-\frac{K_z^2}{K^2}\right)  & \left(\nu + 2(\fsgn)^2\stokes K_x\right)
=\nonumber\\
& -\stokes \fsdn 
\left(\rmi \nu^2
+ 2\fsgn K_x\frac{K_z^2}{K^2}\right)\mathcal{F}(\wstar )\, ,
\label{eqDispersion}
\end{align}
with $\nu=\omega/\Omega$, $K = \eta k/\Omega^2$, $\mathcal{F}(\wstar )=(1+\wstar )(1-\mathcal{I})/\wstar $, and we have defined an average Stokes number 
\begin{align}
\stokes=\Omega\bartausn.
\end{align} 
Note that in the monodisperse limit, the average Stokes number $\stokes$ is equal to the Stokes number of the single sized dust fluid. It is therefore possible to discuss the SI in terms of $\stokes$, remembering that $\stokes \rightarrow \Omega \taus(a^*)$, where $a^*$ is the single dust size under consideration.

\subsection{Power law size distributions}
We focus on power law size distributions, such that $\sigma^0(a) \propto a^{\beta +3}$ between a minimum size and a maximum size. Here $\beta=-3.5$ corresponds to the canonical MRN distribution \citep{1977ApJ...217..425M} representative of the grain size distribution in the interstellar medium. Power law size distributions allow for closed form expressions for $\mathcal{I}$. For example, for $\beta=-3.5$:
\begin{align}
\mathcal{I}(\wstar )
=  1 +
\wstar \left[\frac{\sqrt{w}}{1-\sqrt{s}}\tan^{-1}\left(\frac{1-\sqrt{s}}{\sqrt{w}+\sqrt{s/w}}\right)  - 1\right]\, ,
\label{eq:intI}
\end{align}
where $s=a_{\rm min}/a_{\rm max}$ and $w = \bartausn \wstar /\taus(a_{\rm max})$. 

\subsection[The Long and Short of It]{The Long and Short of It\footnote{In homage to YG05.}}
\label{sec:wavelength}

In this section, we consider three different wave length limits that help us understand the behaviour of the TV-PSI. Readers not interested in technical details may wish to skip to section \ref{sec:summary}, where the results are summarized.

\subsubsection{Growing modes at short radial wave lengths ($\stokes K_x \gg K_z/K$)}
\label{sec:short}

We first consider large radial wave numbers, formally letting $\stokes K_x \gg K_z/K$, and show that we can find exponentially growing modes for the TV-PSI. Note that these radial wave numbers are therefore much larger than where the monodisperse SI has its maximum growth (\citetalias{2005ApJ...620..459Y}, \citealt{2018MNRAS.477.5011S}). If we formally set $K_z/K = O(1)$, we find from (\ref{eqDispersion}) that in this limit
\begin{align}
\nu^2 = \frac{K_z^2}{K^2}\left(1- \mu\mathcal{F}(\wstar )\right),
\label{eq:analimit}
\end{align}
where $\mu$ is the dust to gas ratio. 
For a monodisperse dust fluid, $\mathcal{F}=1$, and we recover the high-$\mu$ SI for $\mu>1$ \citep{2018MNRAS.477.5011S}. In the limit $K_x \stokes \gg 1$, we have that $\wstar  \rightarrow -\fsdn$, so for a polydisperse dust fluid we end up with:
\begin{align}
\nu^2 = \frac{K_z^2}{K^2}\left(2- \mathcal{I}(-\fsdn)\right).
\end{align}
Interestingly, $\mathcal{I}(-f_d^0),$ which is an integral of a real function over a real interval, is found to pick up an imaginary part from the residue of the pole at $\taus(a) = \fsdn\bartausn$. Unlike for the monodisperse SI, this means that in principle growth through this channel is possible for $\mu < 1$ in this wave number regime if the pole is in the integration domain, which is the case if the size distribution is wide enough. For the MRN size distribution, the imaginary part of $\mathcal{I}(-f_d^0)$ is found to be negative, leading to growth according to (\ref{eq:analimit}). We comment that while formally we have assumed $K_z/K=O(1)$, which is valid in the wavenumber regime we study numerically below, the analysis is equivalent if we set $K_z/K = O(\stokes)$, indicating that (\ref{eq:analimit}) is valid for $K_z \sim \stokes K_x \ll K_x$ as well.

\subsubsection{The absence of secular modes at long radial wavelengths ($K_x \ll 1/\stokes$)}
\label{sec:long}

The SI has growing modes for dust to gas ratios $\mu < 1$, where the SI is a true resonant drag instability \citep{2018MNRAS.477.5011S}. These modes were first analyzed in the regime $K_x \sim K_z = O(1)$ \citepalias{2005ApJ...620..459Y}. In this section, we specifically assume $K_x \ll 1/\stokes$ and $K_z \gtrsim K_x$. In this case, we develop a series in $\stokes$ and formally write $\nu = \nu_0 + \stokes \nu_1$. Note that when $\nu = O(1)$ (epicyclic mode), $\wstar  = O(\stokes^{-1})$, while if $\nu = O(\stokes)$ (secular mode), $\wstar  = O(1)$. In the former case, one can show that 
\begin{align*}
\mathcal{F}(\wstar ) = 1 + O(\stokes)\, .
\end{align*}
For the epicyclic mode, we then find
\begin{align}
\nu = \pm K_z/K - \stokes  \fsdn\left(\rmi/2+ \fsgn K_x\right) + O(\stokes^2)\, .
\end{align}
Hence, the epicyclic TV-PSI mode is always damped in this wave length regime. Looking for a secular mode with $\nu = O(\stokes)$, at lowest order we get
\begin{align}
\nu_1 = 2\fsgn K_x \left(\fsdn \mathcal{F}(\wstar _1)-\fsgn\right)\, ,
\label{eq:nu1}
\end{align}
with
\begin{align*}
\wstar _1 = \frac{\nu_1 - 2\fsgn \fsdn K_x}{2\fsgn K_x}\, .
\end{align*}
Note that in the limit of a monodisperse dust fluid, which has $\mathcal{F}=1$, we recover the secular mode of the SI, which shows instability at order $\stokes^3$ (\citetalias{2005ApJ...620..459Y}, \citealt{2011MNRAS.415.3591J}).

In terms of $\wstar _1$ we find that in order for (\ref{eq:nu1}) to be satisfied and therefore to have a secular TV-PSI mode (growing or decaying), we need either $\wstar _1=-1$, which we must discard as this solution was introduced by multiplying both sides by $1+\wstar $ to get to (\ref{eqDispersion}), or:
\begin{align}
\wstar _1  = \fsdn (1-\mathcal{I}(\wstar _1))\, .
\label{eqZ1}
\end{align}
For an MRN size distribution, using the explicit expression for $\mathcal{I}$ (\ref{eq:intI}), we find that we must have:
\begin{align}
\frac{1}{\mu}  = 
-\frac{\sqrt{w}}{1-\sqrt{s}}\tan^{-1}\left(\frac{1-\sqrt{s}}{\sqrt{w}+\sqrt{s/w}}\right)\, .
\end{align}
In the limit $s\rightarrow 0$, the real part of the right hand side is $<0$ for all $w$, which means the secular SI mode does not exist for a wide enough size distribution. We have found this to be qualitatively independent of the power law index $\beta$. 
 Therefore, in the TV approximation, no secular modes (growing or decaying) exist for the PSI at these wave numbers. This means that there is no classical stability boundary in the sense that upon widening the size distribution, one can track a growing SI mode to a decaying PSI mode, crossing zero growth at the stability boundary. The mode simply ceases to exist for wide enough size distributions.

\subsubsection{Analysis at intermediate wave lengths}
\label{sec:rdi}

As was noted already in \citetalias{2005ApJ...620..459Y}, while the regime $K_x \ll 1/\stokes$ is relatively easy to analyze, maximum growth of the SI is achieved when  $K_x \sim 1/\stokes$.  For classical monodisperse SI, this is the regime of RDI wave numbers when $\mu<1$ \citep{2018MNRAS.477.5011S}. It is worth noting that there is no polydisperse RDI theory. In the limit of $\mu \ll 1$, it was shown by \citetalias{2019ApJ...878L..30K} that one obtains a superposition of independent two-fluid instabilities. Here we show that also at RDI wave numbers defined by $\stokes$, which is an average over the size distribution, the TV-PSI at $\mu < 1$ has no growing modes for wide enough size distributions. Due to the more difficult nature of the problem at these wave numbers, this analysis is necessarily less rigorous compared to section \ref{sec:long}.

The low-$\mu$ monodisperse SI is a resonant drag instability \citep{2018ApJ...856L..15S}, which grows fastest if the wavenumbers $K_x$ and $K_z$ are related by \citep{2018MNRAS.477.5011S}:
\begin{align}
4\stokes^2 K_x^2 = \frac{K_z^2}{K^2}(1+\mu)^2.
\label{eq:rdi}
\end{align}
Note that since we are talking about the monodisperse SI, the average Stokes number equals the Stokes number of the single particle size under consideration. While there is no RDI theory for the polydisperse case, we can use (\ref{eq:rdi}) with the average Stokes number as a measure of an intermediate wave length, in between those considered in sections \ref{sec:short} and \ref{sec:long}.
It is worth noting that the relation $K_x^2 \propto K_z^2/K^2$ appears more generally in SI calculations than just the RDI. Specifically, instability regions for $\mu > 1$, where the SI is \emph{not} and RDI \citep{2018MNRAS.477.5011S}, have a similar shape.

At the RDI resonance, we find from the non-dimensional dispersion relation (\ref{eqDispersion}) that
\begin{align}
\left(\nu^2-\hat K_z^2\right)  \left(\nu +\fsgn\hat K_z\right)
=-\fsdn\left(\rmi \stokes \nu^2+ \hat K_z^3\right)\mathcal{F}(\wstar)\, ,
\end{align}
with $\hat K_z = |K_z|/K < 1$ (we focus on $K_z > 0$). One can  develop a series up to first order in the small parameter $\stokes$, writing $\nu = \nu_0 + \stokes\nu_1$:
\begin{align}
\left((\nu_0 + \stokes\nu_1)^2-\hat K_z^2\right)  &\left(\nu_0 + \stokes\nu_1 +
  \fsgn\hat K_z\right)
=\nonumber\\
&-\fsdn\left(\rmi \stokes \nu_0^2+ \hat K_z^3\right)\mathcal{F}(\wstar)\, ,
\label{eq:disp1}
\end{align}
with
\begin{align}
\mathcal{F}(\wstar) =& \mathcal{F} (\wstar_0) +\mathcal{F} '(\wstar_0)(\wstar-\wstar_0) \nonumber\\
=& \mathcal{F} (\wstar_0) +\mathcal{F} '(\wstar_0)\stokes \wstar_1.
\end{align}
At lowest order in $\stokes$, (\ref{eq:disp1}) reads:
\begin{align}
\left(\nu_0^2-\hat K_z^2\right)  \left(\nu_0 +
  \fsgn\hat K_z\right)
=-\fsdn \hat K_z^3\mathcal{F}(\wstar_0)\, ,
\label{eq:zerothorder}
\end{align}
while at first order we get that
\begin{align}
\left(3\nu_0^2 + 2\nu_0\fsgn\hat K_z - \hat K_z^2+ \fsdn \hat K_z^2\mathcal{F}'(\wstar_0)\right)\nu_1 
=\nonumber\\
\fsdn\left(\fsdn \hat K_z^3\mathcal{F}'(\wstar_0)-\rmi \nu_0^2\mathcal{F}(\wstar_0)\right)
\label{eq:firstorder}\, .
\end{align}
For the low-$\mu$ SI, we are looking for growth rates of order $\stokes$ at the resonant wave numbers (\citealt{2018MNRAS.477.5011S}, \citetalias{ 2005ApJ...620..459Y}). For a monodisperse dust component, $\mathcal{F}=1$, so that (\ref{eq:zerothorder}) gives the mode oscillation frequency ($\nu_0$ is real), while (\ref{eq:firstorder}) determines the growth rate ($\nu_1$ is imaginary). While the solution of the cubic in (\ref{eq:zerothorder}) is unwieldy, simple expressions can be derived by considering $\fsdn \ll 1$ \citep{2018MNRAS.477.5011S}. For the TV-PSI, we face the additional difficulty that $\mathcal{F}$ involves an integral. Nevertheless, by inspecting the function $\mathcal{F}$ we can provide some approximate bounds on the width of the size distribution that allows growing modes in this regime.

\begin{figure}
	\includegraphics[width=0.9\columnwidth]{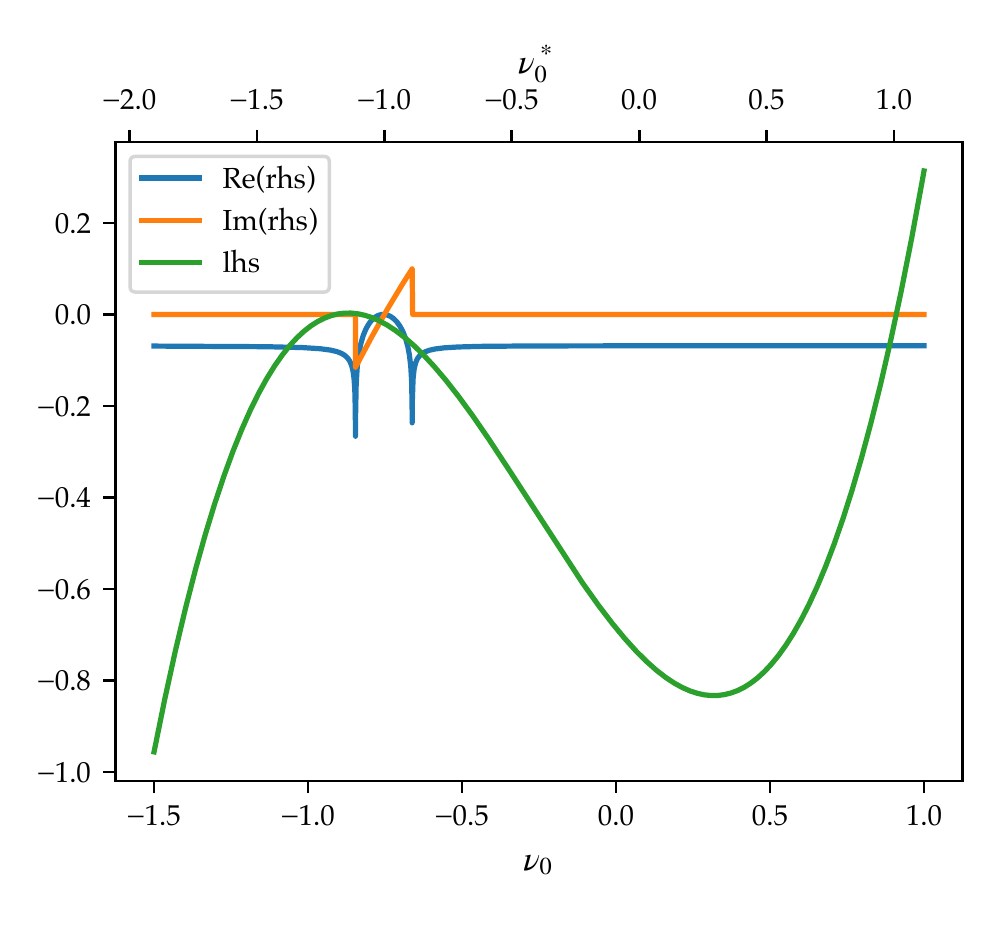}
    \caption{The left and right hand sides of the zeroth order dispersion relation (\ref{eq:zerothorder}), for $\mu=0.1$ and $\hat K_z=0.91$ and an MRN size distribution with stopping time range $\Omega\taus \in [0.008, 0.01]$.}
    \label{fig:existence}
\end{figure}

\begin{figure}
	\includegraphics[width=0.9\columnwidth]{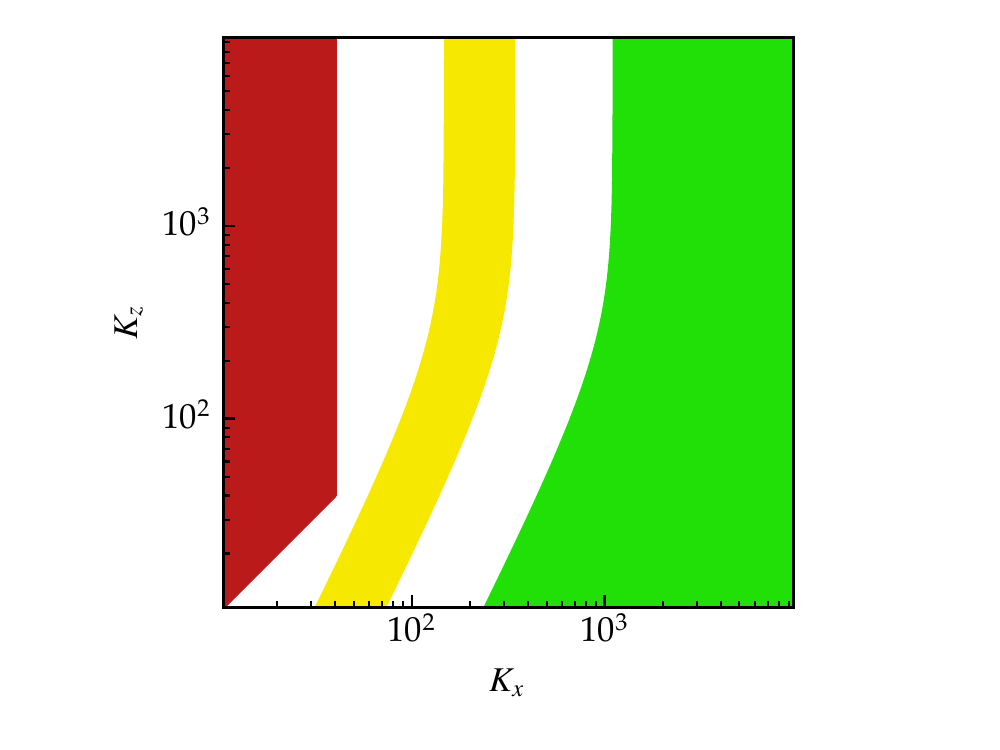}
    \caption{Schematic overview of the wave number ranges considered in section \ref{sec:wavelength}. In the red region, no secular modes exist, see section \ref{sec:long}. In the yellow region (RDI wave numbers), no growing modes exist for $\mu < 1$, see section \ref{sec:rdi}. In the green region, growing PSI modes were found, see section \ref{sec:short}. Specific parameters used for this plot were $\mu=0.5$, $10^{-8} < \Omega \taus < 10^{-2}$, and therefore $\stokes=0.0034$.}
    \label{fig:overview}
\end{figure}

In Figure \ref{fig:existence} we show the left hand side and right hand side of (\ref{eq:zerothorder}) as a function of $\nu_0$. The left hand side is a cubic, while the right hand side is a constant times $\mathcal{F}$. 
The region around $\wstar_0=-1$ where $\mathcal{F}$ has an imaginary part denotes the region where the resonance $\wstar_0 = -\taus(a)/\bartausn$ is inside the integration domain. It is therefore bound by
\begin{align}
-\frac{\tau_{\rm s, max}}{\bartausn} <\wstar_0 < -\frac{\tau_{\rm s, min}}{\bartausn},
\end{align}
or, in terms of $\nu_0$ and $\stokes$:
\begin{align}
-\frac{\mathrm{St}_{\rm max}}{\stokes} <  \frac{\nu_0 - \fsdn \hat K_z}{\hat K_z} < -\frac{\mathrm{St}_{\rm min}}{\stokes}.
\end{align}
Roots, whenever they are located in this region are always found to be either neutral or damped. This is consistent with the observation that growth in this region of parameter space is first order in Stokes number. Note that in this instance, the size resonance does not promote growth. The two main differences between the current situation and that of section \ref{sec:short} are (i) $\mathcal{F}$ now depends on $\nu$ (which leads to a change in sign of the imaginary part, as is clear from Figure \ref{fig:existence}) and (ii) the dispersion relation is different, so that a positive imaginary part of $\mathcal{F}$ no longer directly translates into a growing mode. The growing mode is therefore found to exist if it is located to the right of the `imaginary region' as seen in Figure \ref{fig:existence}. That is, for growth we need
\begin{align}
\nu_0  > -\hat K_z\left(\frac{\mathrm{St}_{\rm min}}{\stokes} + \fsdn\right).
\label{eq:limit}
\end{align}
A simple estimate for when this happens can be obtained by approximating $\mathcal{F}$ by unity outside the 'imaginary region'. This means that $\nu_0$ is found from the cubic
\begin{align}
\left(\nu_0^2-\hat K_z^2\right)  \left(\nu_0 +
  \fsgn\hat K_z\right)
=-\fsdn\hat K_z^3\, .
\end{align}
Roots of cubic equations are unwieldy. A simple estimate can be obtained by considering the limit of small $\fsdn$ \citep{2018MNRAS.477.5011S}. First, for ease of notation define $X = \nu_0/\hat K_z$ and $\delta=\fsdn$:
\begin{align}
\left(X^2-1\right)  \left(X + 1-\delta\right) + \delta = 0 \, .
\label{eq:Xdisp}
\end{align}
Look for a solution of the form $X = X_0 + \delta^{1/2} X_1$:
\begin{align}
\left((X_0 + \delta^{1/2} X_1)^2-1\right)  \left(X_0 + \delta^{1/2} X_1 + 1-\delta\right) + \delta = 0 \, .
\end{align}
At lowest order we find for the relevant mode $X_0 = -1$. At order $\delta^1$ we find that
\begin{align}
- 2X_1^2 + \delta = 0 \, ,
\end{align}
and hence that $X_1 = \pm 1/\sqrt{2}$. The relevant root can then be approximated by $X = -1 + \sqrt{\fsdn/2}$, or
$\nu_0= -\hat K_z + \hat K_z\sqrt{\fsdn/2}$. Note that a similar approximation in terms of dust to gas ratio was found by \cite{2018MNRAS.477.5011S}. There is no value in going to higher orders in $\delta$ since the main error now stems from the approximation that $\mathcal{F}$ is constant. Using this approximation for $\nu_0$ in (\ref{eq:limit}) we find that in order to have a growing mode we need
\begin{align}
\frac{\mathrm{St}_{\rm min}}{\stokes} \gtrsim \frac{1 - \sqrt{\fsdn/2}}{1-\fsdn} - \fsdn.
\end{align}
For the parameters of the runs depicted in Figure \ref{fig:existence}, this estimate predicts growth is possible for $\mathrm{St}_{\rm min}/\stokes \gtrsim 0.77$. Numerically, we find growth is possible for $\mathrm{St}_{\rm min}/\stokes > 0.68$.  Therefore, for wide enough size distributions there are no growing TV-PSI modes at RDI wave numbers for $\mu < 1$.  This is indicated by the yellow region in figure \ref{fig:overview}. Below, we show that there are also no growing modes in a finite region around the exact RDI wavenumbers. It is worth noting that, similar to the long wavelength regime of section \ref{sec:long}, there is no classical stability boundary where the mode crosses the zero growth line: widening the size distribution leads to the mode disappearing altogether. 

While only at $\mu <1$ is the SI an RDI, it is worth briefly considering a similar wave number range for $\mu > 1$, as the basic relation between $K_x$ and $K_z$ (\ref{eq:rdi}) is still relevant for $\mu > 1$ as well as the TV-PSI. In particular, consider the wave number relation
\begin{align}
4 a^2\stokes^2K_x^2 = \hat K_z^2 (1+\mu)^2,
\label{eq:aRDI}
\end{align}
which has the effect of shifting the wave numbers to higher $K_x$ compared to RDI wave numbers for $a < 1$\footnote{Note that we require $a=O(1)$ in order for the ordering in $\stokes$ to remain valid.}. We can then ask whether any growing modes at order $\stokes^0$ exist for a given value of $\mu$. The dispersion relation at zeroth order (\ref{eq:Xdisp}) now reads 
\begin{align}
b^2Y^3+  b^2Y^2 - Y  - 1  + \mu = 0 \, ,
\end{align}
with $Y = X/\fsgn$ and $b=a \fsgn$. The discriminant $\Delta$ of this cubic signals the transition between three real roots ($\Delta > 0$) and one real root and two complex conjugate roots ($\Delta < 0$). In the latter case, one of the complex roots is growing, and for the monodisperse case we have entered the domain of the high-$\mu$ SI \citep{2018MNRAS.477.5011S}. We have that $\Delta =0$ when
\begin{align}
4b^4(\mu-1) + \left[18(\mu-1)  - 1 +27(\mu-1)^2\right]b^2 - 4 = 0\, .
\label{eq:bRDI}
\end{align}
It is straightforward to see that real solutions for $b$ exist when $\mu>1$, the high-$\mu$ SI. For higher radial wave numbers than the boundary determined by (\ref{eq:aRDI}), growth rates at order $\stokes^0$ exist for the SI in the TV approximation. Since for the relevant root we can approximate $\mathcal{F}=1$ (as was done above), we expect the high-$\mu$ TV-PSI to give growing modes at order $\stokes^0$ for similar wave number combinations as the high-$\mu$ SI. Note, however, that while the SI has growing modes at order $\stokes$ for radial wave numbers smaller than given by (\ref{eq:aRDI}), for the high-$\mu$ TV-PSI we expect an abrupt transition between fast growth and no growth at all. 

As a corollary, there are no real solutions for $b$ when $\mu<1$. This means that it is not possible in this case to shift the RDI curve so that it falls in a region with growing modes in this wavelength regime. Therefore, for $\mu<1$ not only do we not get any growing modes at the exact RDI wavenumbers (defined by the average Stokes number), but there are also no growing modes to be found in the vicinity. For smaller values of $\mu$, the region of growing modes as identified in section \ref{sec:short} will be further away from the RDI wavenumbers based on $\stokes$.

\subsubsection{Summary}
\label{sec:summary}

In Figure \ref{fig:overview} we show schematically the three wave number regimes considered previously in this section. The wave number ranges are chosen to match those of the numerical simulations presented in section \ref{sec:numerical}, and for definitiveness we have taken $\mu=0.5$, $10^{-8} < \Omega \taus < 10^{-2}$, and therefore $\stokes=0.00334$ to calculate the position of the RDI wavenumbers, again in order to match the numerical results below. While in the TV approximation, the SI has growing modes for all $K_x$ and $K_z$ in the range depicted in Figure \ref{fig:overview}, the TV-PSI only has growing modes in the green region for all values of $\mu$. In the red region, defined by $K_x \ll 1/\stokes$ and $K_z \gtrsim K_x$, the secular mode responsible for SI growth as identified by \citetalias{2005ApJ...620..459Y} and further studied by \cite{2011MNRAS.415.3591J} does not exist for wide enough size distributions (section \ref{sec:long}). We stress again that this mode does not cross the zero growth line and becomes damped: it ceases to exist. In the yellow region, centered around the RDI wave number range, the SI grows fastest, but the TV-PSI has no growing modes for sufficiently wide size distributions for $\mu < 1$ (section \ref{sec:rdi}). For $\mu > 1$, both SI and TV-PSI have growing modes in the yellow region (section \ref{sec:rdi}). The green region, defined by $\stokes K_x \gg K_z/K$, is where growing TV-PSI modes can occur for all values of $\mu$ (section \ref{sec:short}). We should stress that this section deals exclusively with modes that exist under the TV approximation. When solving the full system in section \ref{sec:numerical}, additional growing modes show in the red region but with growth rates a few orders of magnitude lower than those found in the green region.

\begin{figure}
	\includegraphics[width=0.9\columnwidth]{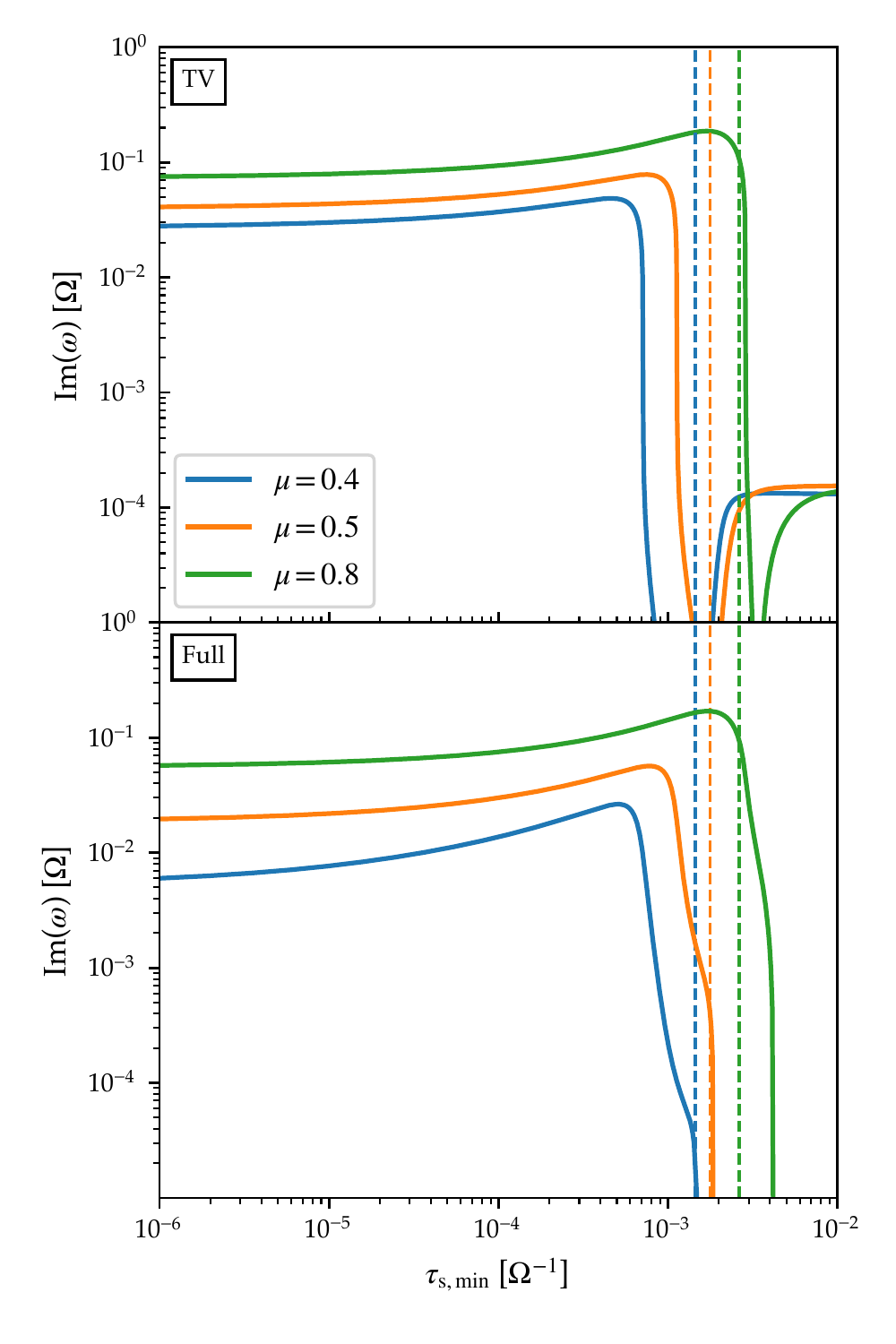}
    \caption{ Demonstration of the onset of the size resonance in PSI. {\sl Top:} Growth rates for the TV limit of PSI, using an MRN size distribution in stopping time range $[\tau_\mathrm{s, min}, 0.01\Omega^{-1}]$ for three different dust to gas ratios, all at $K_x=10^{2.75}$,  $K_z=500$. The vertical dashed lines indicate the position of the resonance.
    {\sl Bottom:} Results for the same parameters in the full PSI eigenproblem. Note that the monodisperse SI for $\mu=0.4$, $\Omega\taus=10^{-2}$ and $K_z=500$ is stable when $K_x>10^{2.29}$, as is the case for the monodisperse limit of these plots.}
    \label{fig:size}
\end{figure}

\begin{table*}
	\centering
	\caption{Numerical results for an MRN size distribution with $\Omega\taus = [10^{-4}, 10^{-1}]$ and $\mu=1$. Wave numbers and mode frequency $\nu_{\rm K+19}$ were produced from digitized versions of Figure 1 and 2 of \citetalias{2019ApJ...878L..30K}. $\nu$ is the result from our eigensolver, while $\nu_{\rm TV}$ is the result from our terminal velocity solver.}
	\label{tab:krapp}
	\begin{tabular}{ccccc} 
		\hline
		$K_x$ & $K_z$ & $\nu$ & $\nu_{\rm K+19}$ & $\nu_{\rm TV}$\\
		\hline
		 $20.6819$ & $6.40829$
& $0.221331+0.015918\rmi$ & $0.227234+0.0161574\rmi$ & $0.22147394+0.03665379\rmi$\\
		\hline
	\end{tabular}
\end{table*}

\subsection{SI versus TV-PSI}

In this section, we highlight the main differences between the monodisperse streaming instability and the PSI in the terminal velocity approximation.

{\bf SI:} The three wavelength regimes discussed in the section \ref{sec:wavelength} highlight the different character of the SI for dust to gas ratios $\mu<1$ compared to $\mu > 1$. In the former case, growth at order $\stokes$ is possible at the RDI resonant wavenumber \citep{2018MNRAS.477.5011S}. For the high-$\mu$ SI, on the other hand, much larger growth rates are observed that are independent of $\stokes$ \citep{2018MNRAS.477.5011S,2020MNRAS.tmp.2397S}. In the thermodynamic \citep{2017ApJ...849..129L} interpretation of the one fluid model, for $\mu > 1$ cooling $\mathcal{\hat C}_{\rm g}$ and density perturbations are exactly in phase leading to instability, while for $\mu<1$, the cooling perturbation is exactly out of phase with the density perturbation, leading to stable epicyclic oscillations and growth at higher order \citep{2018MNRAS.477.5011S}. 

\noindent{\bf Short wave length TV-PSI:} While for the SI in the TV approximation there is a sharp dividing line for fast growth at $\mu=1$ (equation (\ref{eq:analimit}) with $\mathcal{F}=1$), for the TV-PSI  at wavelengths much shorter than the RDI wave length based on $\stokes$ there is \emph{always} a phase difference between density and cooling if the resonance condition $\wstar  = -\taus(a)/\bartausn$ is met, which in the short wave length limit translates into $\taus(a)=\fsdn \bartausn$ (see section \ref{sec:short}). The resonance triggers a strong response at a specific $\taus$ that is different from $\bartausn$, which therefore leads to a phase difference in cooling (contribution from resonant size) and density (contribution from $\bartausn$). Growth at order $\stokes^0$ is possible only for a size distribution that is wide enough to include the resonance. This is illustrated in  the top panel of Figure \ref{fig:size}, where in the single size limit (right-hand side of the figure), only slow growth of the secular mode is found because $\mu < 1$. If the minimum stopping time is decreased so that the resonance (vertical dashed lines) is included, growth rates are substantial. Thus, it appears that the instability abruptly changes character when the size resonance is included. Intuitively this makes sense: a formally infinite perturbation at a specific size is communicated to the gas through an integral over size, which in turn makes all sizes feel the effect of the resonance. We note that \citetalias{2019ApJ...878L..30K} also found faster growth for size distributions compared to the monodisperse limit in some cases. The bottom panel of Figure \ref{fig:size} shows that this feature is not specific to the TV approximation: the full PSI model also shows an abrupt rise in growth rates once the size resonance is inside the integration domain. Note that the full model has no growth in the single size limit at these large wave numbers; a well-known characteristic of the monodisperse SI (e.g. \citetalias{2005ApJ...620..459Y}, \citealt{2007ApJ...662..613Y}).

\noindent{\bf Long wave length TV-PSI:} In the long radial wavelength regime (much longer than RDI wave lengths), the lack of growing modes for wide size distributions is due to the disappearance of the secular mode (section \ref{sec:long}). This mode was interpreted in the monodisperse  case by \cite{2011MNRAS.415.3591J} as due to a buoyancy force arising in two-fluid systems. We find that a polydisperse dust fluid does not seem to be able to provide a coherent buoyancy restoring force, which means the SI secular mode is effectively replaced by the neutral mode $\wstar =0$ for wide enough size distributions.  At intermediate wave lengths for $\mu<1$, where the SI has its maximum growth at the RDI resonance, the TV-PSI has no growing modes (section \ref{sec:rdi}).

\begin{figure*}
	\includegraphics[width=0.9\textwidth]{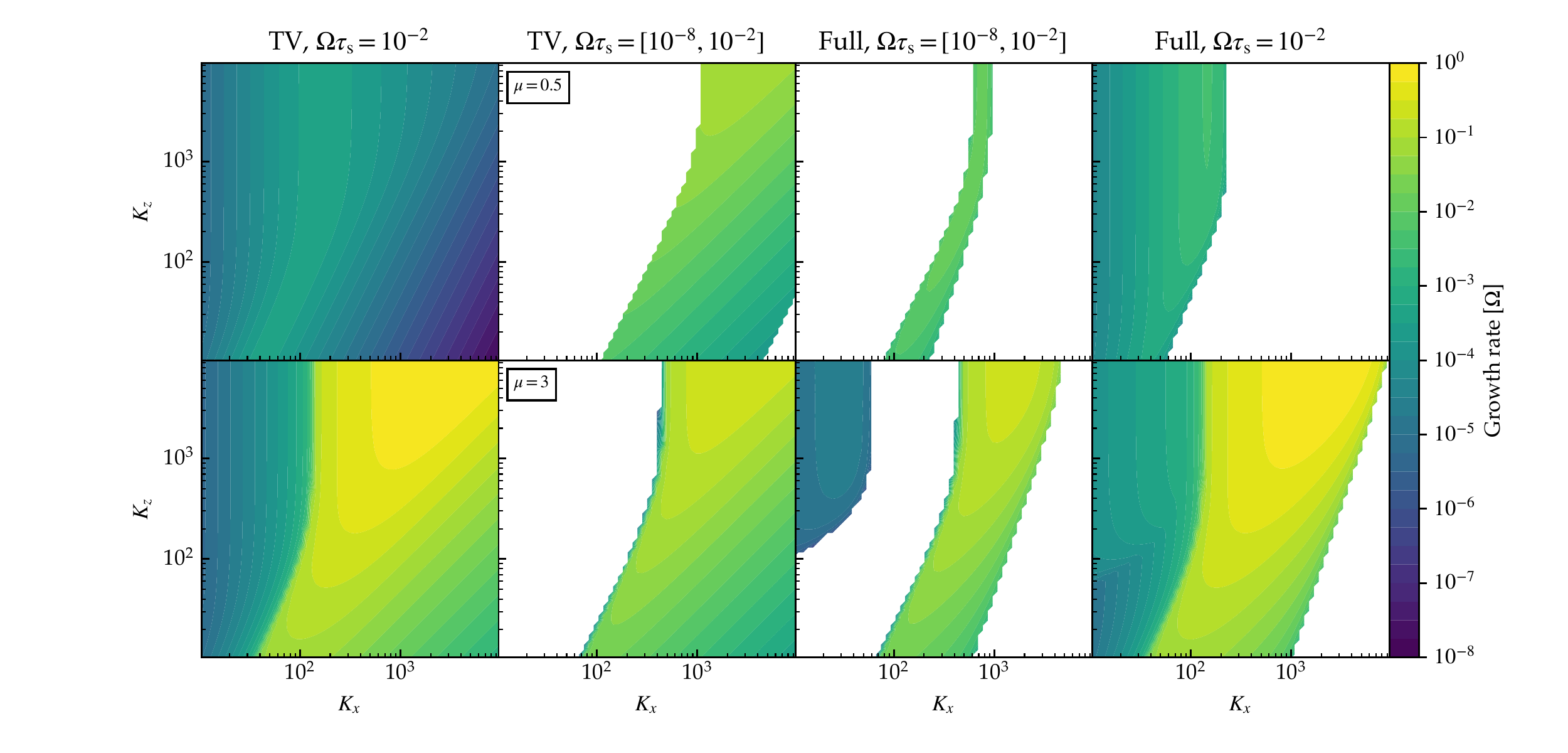}
    \caption{Growth rates (in units of $\Omega$) of the PSI with an MRN size distribution for $10^{-8} < \Omega\tau_s < 10^{-2}$ compared to the SI with $\Omega\tau_s=10^{-2}$. From left to right: TV monodisperse SI, TV-PSI, full model PSI, full model  monodisperse SI. Top panels: $\mu = 0.5$, bottom panels: $\mu=3$.}
    \label{fig:kxkz}
\end{figure*}

\section{Numerical results}
\label{sec:numerical}
We solve the TV dispersion relation (\ref{eqDispersion}) using Newton's method in the complex plane, using the fastest growing mode in the single size case as an initial guess. We solve the full eigenvalue problem (\ref{eqFullFirst})--(\ref{eqFullLast}) by discretizing (\ref{eqFullSigma})--(\ref{eqFullLast}) in $\taus(a)$ by collocation on a Chebyshev grid, and computing the integral terms in (\ref{eqFullVg}) via trapezoid rule. The resulting matrix representation of the eigenproblem is solved with the Python {\tt numpy.linalg.eig} routine. Details on the numerical method are presented in Appendix \ref{app:numerical}.

While the TV solver can treat a size continuum without the need for any discretization, the eigenvalue solver uses numerical quadrature to compute the integral terms in the perturbation equations. It is important to contrast our numerical approach to the eigenvalue problem with that of \citetalias{2019ApJ...878L..30K}. Both numerical methods aim at studying a size continuum, but have different ways of getting there. One important difference is that our equilibrium state is always set by the continuous size distribution (see (\ref{eq:veqfirst})-(\ref{eq:veqlast})), independent of the number of collocation points used in discretizing (\ref{eqFullSigma})--(\ref{eqFullLast}). In contrast, in \citetalias{2019ApJ...878L..30K} the equilibrium state depends on the number of dust fluids considered \citep[see][in particular their equations (79)-(82)]{2019ApJS..241...25B}. This means that while both methods should give similar\footnote{Similar, not exactly equal because \citetalias{2019ApJ...878L..30K} use a different drag law, with a stopping time that is independent of gas density. Since gas density variations are typically very small, the resulting difference should be small \citep{2007ApJ...662..613Y}.} results in the limit of an infinite number of dust fluids (in the case of \citetalias{2019ApJ...878L..30K}) and an infinite number of collocation points (in our case), at finite resolution differences can be expected \emph{because the underlying physical model is different}. At finite resolution, \citetalias{2019ApJ...878L..30K} compute the \emph{exact} growth rates for a \emph{finite} set of single-sized fluids, while we compute the \emph{approximate} growth rates for a \emph{continuous} size distribution. Meaningful comparisons can therefore only be made in the continuum limit, which means that for comparisons we are limited to results of \citetalias{2019ApJ...878L..30K} that converge with number of dust fluids\footnote{A lack of convergence in the continuum limit would also be a meaningful comparison, but is hard to obtain in practice.}. They present one such case in their Figure 2. We have extracted the relevant wave numbers from a digitized version of their Figure 1 (the position of the white triangle in the top middle panel) and compare our result to their converged result presented in the orange curve of their Figure 2 (again from a digitized version) in Table \ref{tab:krapp}. The eigenvalues agree to within $3\%$. Further details and more comparisons are presented in Appendix \ref{app:krapp}. We also quote the result of the TV solver in Table \ref{tab:krapp}, which yields a growth rate that is a factor of 2 too large. This could signal that the maximum Stokes number considered in the size distribution ($\mathrm{St}_{\rm max}=0.1$) is too large for the TV approximation to apply, but as we will see below the TV approximation also breaks down for smaller Stokes numbers towards large $K_x$ \citep[as was also observed in][]{2018MNRAS.477.5011S}.  

Our main results are displayed in Figure \ref{fig:kxkz}, where we compare the PSI to the SI for $\mu=0.5$ (top panels) and $\mu=3$ (bottom panels). We note that for the PSI results, an MRN size distribution with $10^{-8} < \Omega \tau_s < 10^{-2}$ gives $\stokes=0.00334$. While in the TV equations, the gas sound speed does not appear, for the full solver we have used the canonical value of $\eta/(c\Omega) = 0.05$ \citep[e.g.][]{2007ApJ...662..613Y}. The gas-dominated SI (top left for TV, top right for full two-fluid) has to rely mostly on the secular mode, yielding growth rates of $\sim \stokes$. The PSI (top middle panels: left for TV, right for the full model) shows larger growth rates for $\stokes K_x \gg K_z/K$, which is the short wavelength limit of section \ref{sec:short}. We found these growth rates to be independent of maximum stopping time, but going to smaller values of $\mu$ pushes these modes to even larger wave numbers. For smaller $K_x$, no growing modes were found, as the secular SI mode does not exist for this wide size distribution (see section \ref{sec:long}). In particular, we note that at the RDI wave numbers  for $\stokes=0.00334$, which for $K_z \rightarrow \infty$ has $K_x \approx 200$ the PSI was found to have no growing modes, consistent with the analysis of section \ref{sec:rdi}. This is further explored in Figure \ref{fig:tvcompare} below. For vertical wave numbers smaller than shown in Figure \ref{fig:kxkz}, TV PSI growth rates in the unstable band steadily decrease. 

The dust-dominated regime (bottom panels of Figure \ref{fig:kxkz}) shows large growth rates in the SI limit (bottom left and right), which is the high-$\mu$ SI. For the PSI, growth rates are moderately reduced but again found to be confined to large wave numbers. The full model PSI (lower row, middle right) displays an island of very small growth rates along the $K_z$ axis.  This is due to instability of the epicyclic mode, which was dropped from the TV model at these wave numbers. Towards smaller $K_x$, outside the range of Figure \ref{fig:kxkz}, these growth rates decay until they drop below $2\cdot 10^{-7}$ at $K_x\sim 0.1$. In both gas and dust dominated regimes, although the TV approximation  predicts the low $K_x$ cutoff of instability well, 
it does not capture the high-$K_x$ cutoff for growth rates found in the full model. For the SI, this is known to be due to the neglect of higher order terms in $\stokes$ in TV. The wider the instability strip in the full model, the better agreement with the analytic limit (\ref{eq:analimit}), varying from $\sim 20\%$ in the bottom panels to a factor of $\sim 3$ in the very narrow instability strip in the upper panels. Further discussion on the validity of TV is presented in Appendix \ref{app:tv_valid}.

\begin{figure}
    \begin{center}
	\includegraphics[width=0.7\columnwidth]{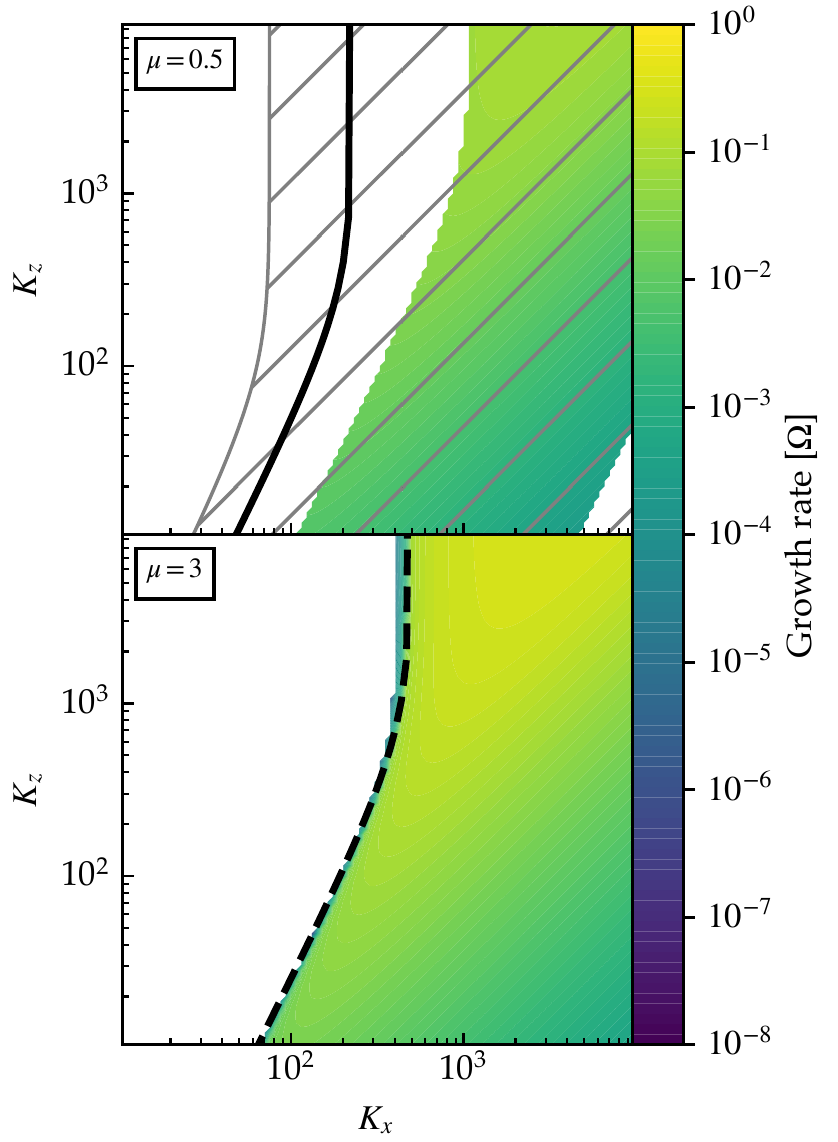}
    \end{center}
    \caption{ Growth rates (in units of $\Omega$) of the TV-PSI with an MRN size distribution for $10^{-8} < \Omega\tau_s < 10^{-2}$. {\sl Top panel}: $\mu = 0.5$, {\sl solid black}: RDI resonance condition for $\stokes$ (section \ref{sec:rdi}), demonstrating no growth at the RDI condition applied to the mean Stokes number, {\sl hatched region}: wavenumber range where any single size in the size distribution would have its RDI.
    {\sl Bottom panel}: $\mu=3$, {\sl dashed black}: limit for TV-PSI growth for $\mu>1$ (section \ref{sec:rdi}), showing agreement with the onset of fast growth.}
    \label{fig:tvcompare}
\end{figure}

We compare the predictions of section \ref{sec:wavelength} with the numerical TV-PSI results in Figure \ref{fig:tvcompare}. Note that these are the same panels as shown in the left middle panels of Figure \ref{fig:kxkz}, but with the RDI resonance condition (\ref{eq:rdi}) shown in the black curve in the top panel, while the limit for TV-PSI growth for $\mu>1$ as obtained from (\ref{eq:aRDI}) and (\ref{eq:bRDI}) is shown with the dashed black curve in the bottom panel. From the top panel, it is clear that we did not find any growing TV-PSI modes at RDI wave numbers, in agreement with section \ref{sec:rdi}. We note that when $\mu$ approaches unity, the unstable region and the black curve move closer together. This is consistent with our findings in section \ref{sec:rdi}, since we took the limit $\mu \ll 1$ in order to show that there are no growing modes at RDI wave numbers. The dividing line for having growth at these wave numbers is $\mu \approx 1$. In addition, we also did not find any growing modes at smaller radial wave numbers, in agreement with section \ref{sec:long}. The hatched region indicates the wave number range where any single size in the size distribution would have its RDI. Since we consider six orders of magnitude in stopping time, this region extends beyond the edge of the plot on the right hand side. 

In the bottom panel of Figure \ref{fig:tvcompare} we revisit the case $\mu=3$ of Figure \ref{fig:kxkz}. 
The boundary of the unstable region as found from (\ref{eq:aRDI}) and (\ref{eq:bRDI}), depicted by the dashed black curve, matches the numerical boundary quite well. We note again that this is not a stability boundary in the sense that to the left of the boundary the relevant mode becomes damped: the mode ceases to exist. 
It is worth noting that while the boundary of the unstable region has a similar shape to the RDI curve shown in the top panel, the specific form of the boundary is merely an indication that the TV-PSI involves unstable epicycles, as does the monodisperse SI (\citetalias{2005ApJ...620..459Y}, \citealt{2018MNRAS.477.5011S}, \citealt{2020MNRAS.492.4591J}). Finally, we note that while we only show TV results in Figure \ref{fig:tvcompare}, the TV approximation and the full model always agree very well on the left edge of this instability region, as is clear from Figure~\ref{fig:kxkz}.

\section{Discussion and conclusions}
\label{sec:disc}

We have presented the first analysis of the PSI, which is a version of the SI with a dust component that is a true continuum of sizes. We have focused on the $\stokes \ll 1$ regime  in the TV approximation and MRN dust distributions, and found that exponentially growing modes exist but are confined to very high radial wave numbers ($\stokes K_x \gg K_z/K$) for $\mu \ll 1$. For $\mu>1$, the TV-PSI shares its wavenumber regime where fastest growth (on a dynamical time scale, $O(\Omega)$) occurs with the high-$\mu$ SI. Our analysis was conducted through application of the TV approximation to the continuum equations yielding a simple scalar eigenproblem, and checked by discretizing the unapproximated equations and then conducting a convergence study on the resulting eigenproblem. 
The high wave number nature of the PSI regime explored in this paper ($\stokes K_x \gg K_z/K$, which, for $\eta \sim rh^2\Omega^2$  and $K_z/K=O(1)$, where $h$ is the aspect ratio of the disc and $r$ the fiducial orbital radius of the shearing box, translates into $r k_x \gg h^{-2}\stokes^{-1}$) raises two concerns. First, at sufficiently high wavenumbers the total masses of solids involved in the growing mode become small, which could in turn impact the sizes of resulting solid clumps, ostensibly the seeds of planet formation, in the nonlinear outcome. Second, the dissipative effects of turbulence and particle diffusion more easily damp instability at high wavenumber (\citetalias{2005ApJ...620..459Y}, \citealt{2020ApJ...895....4U}, \citealt{2020ApJ...891..132C}). In addition, turbulence is driven by, and strongly affects the nonlinear phase and planetesimal formation \citep{2007Natur.448.1022J,2011A&A...529A..62J,
    2018ApJ...868...27Y,2020arXiv200110000G}. We will study the effect of turbulence on the PSI in a forthcoming paper.

We have limited the scope of this paper to the MRN dust distribution and $\stokes\ll 1$, and it served well for elucidating the fundamentals of the PSI. The power law slope is well motivated by observations of the ISM, but is likely less appropriate for the midplane regions of an evolved protoplanetary disc where planet formation occurs. At the same time, the SI is often invoked in planet formation models with $\stokes\sim  10^{-1}$--$1$. These conditions exceed the validity of TV. In an upcoming paper we will release these restrictions.

The high-$\mu$ SI is likely the most relevant to the most common scenario for  interest in the SI, that of planetesimal formation \citep{2020MNRAS.tmp.2397S}. The similarity of unstable parameters and growth rates of the PSI gives hope of similar outcomes.

In conclusion, generalising monodisperse SI to include a continuous dust distribution in the form of PSI changes the parameters for instability. For tightly coupled particles to which the TV approximation applies, growth is only possible for $\mu \ll 1$ at radial wavenumbers that are a factor $1/\stokes$ larger than where the SI has its maximum growth.  For $\mu>1$, the PSI has maximum growth rates comparable to the high-$\mu$ SI at similar wave numbers. At these large radial wave numbers, growth time scales can be of the order of a dynamical time scale, while at smaller radial wave numbers, modes that are not part of the TV approximation were found that grow on $\sim 10^4$ dynamical time scales.  

\section*{Acknowledgements}
We thank Min-Kai Lin for sharing benchmark results and Richard Nelson for useful conversations.
This research was supported by an STFC Consolidated grant awarded to the QMUL Astronomy Unit 2017--2020 ST/P000592/1.
This research utilised Queen Mary's Apocrita HPC facility, supported by QMUL Research-IT \citep{apocrita}.
SJP is supported by a Royal Society URF.

\section*{Data availability}

Data available on request.

\bibliographystyle{mnras}
\bibliography{DustSizeContinuum}

\appendix

\section{PSI in the terminal velocity approximation}
\label{app:tv}

\subsection{Polydisperse terminal velocity approximation}
\label{app:tv_derive}

The TV approximation is widely used to study monodisperse gas-dust mixtures in the limit where the coupling is strong. In this context, it first appeared in \cite{2005ApJ...620..459Y} as a tool to study the behaviour of the SI for well-coupled particles. Subsequently, it was extended to a full nonlinear system of evolution equations that could be studied in their own right \citep{2014MNRAS.440.2136L, 2017ApJ...849..129L}. It makes the analysis of well-coupled two-fluid systems considerably easier, both analytically (\citet{2005ApJ...620..459Y}, \citealt{2017ApJ...849..129L}) as well as numerically \citep{2014MNRAS.440.2136L}. Fortunately, the polydisperse equations allow for a similar approximation. 

The full governing equations consist of gas and dust continuity and momentum equations:
\begin{align}
\partial_t\rhog +& \nabla\cdot(\rhog \vg) = 0\, ,\label{eqGasContApp}\\
\partial_t \vg + & (\vg \cdot \nabla)\vg =2\eta {\bf \hat x}\nonumber\\
&  -\frac{\nabla p}{\rhog}-  2\bm{\Omega}\times \vg-\nabla\Phi + \frac{1}{\rhog}\int\sigma \frac{\ud-\vg}{\taus(a)} \rmd a 
\label{eqGasMomApp}\, ,\\
\partial_t\sigma + &\nabla\cdot(\sigma {\bf u} )= 0\, ,\label{eqDustContApp}\\
\partial_t\ud +& (\ud\cdot\nabla) \ud =
-2\bm{\Omega}\times \ud-\nabla\Phi - \frac{\ud-\vg}{\taus(a)}\, .\label{eqDustMomApp}
\end{align}
Using the fact that the size density and size momentum integrate to the dust density and dust momentum:
\begin{align}
\rhod =& \int \sigma(a)\rmd a, \\
\rhod\vd =& \int \sigma(a)\ud (a) \rmd a,
\end{align}
we obtain the size-integrated dust continuity and momentum equations by integrating over dust size $a$:
\begin{align}
\partial_t\rhod +& \nabla\cdot(\rhod \vd) = 0,\label{eqDustContInt}\\
\partial_t\vd +& (\vd \cdot\nabla)\vd = -2\bm{\Omega}\times \vd \nonumber\\
&-\nabla\Phi - \frac{1}{\rhod}\int\sigma\frac{\ud-\vg}{\tau_s(a)}\rmd a
- \frac{1}{\rhod}\nabla\cdot \mathsf{S},
\label{eqDustMomInt}
\end{align}
with stress tensor
\begin{align}
\mathsf{S} = \int\sigma \ud \ud \rmd a-\rhod\vd\vd
= \int \sigma (\ud - \vd)(\ud+\vd)\rmd a. 
\end{align}
In the limit of a monodisperse dust fluid with $\sigma=\rhod \delta(a - a_0)$, where $\delta$ is the Dirac delta function, the integrals become trivial and the stress tensor vanishes, leaving us with the usual dust fluid equations for a single size $a_0$.
 
Subtract the gas momentum equation (\ref{eqGasMomApp}) from the dust momentum equation (\ref{eqDustMomApp}) to obtain an evolution equation for the size-dependent relative velocity between gas and dust: 
\begin{align}
\partial_t\Delta\ud
+ & (\ud\cdot\nabla)\ud
-(\vg\cdot \nabla)\vg =\frac{\nabla p}{\rhog} -2\eta{\bf \hat x}
\nonumber\\
& 
-2 \bm{\Omega} \times\Delta\ud
- \frac{\Delta\ud}{\taus(a)} + 
\frac{1}{\rhog}\int\sigma \frac{\Delta\ud}{\taus(a)} \rmd a.
\end{align}
In the TV approximation, we assume $\Delta\ud = O(\stokes)$ and keep only the lowest order contribution in $\stokes$, so that drag forces adjust quasi-statically to pressure forces \citep{2005ApJ...620..459Y}:
\begin{align}
 \Delta\ud(a) = \taus(a)&\left(\frac{\nabla p}{\rhog}-2\eta{\bf \hat x}\right) \nonumber\\
&+ \frac{\taus(a)}{\rhog}\int\sigma(r) \frac{\Delta\ud(r)}{\taus(r)} \rmd r,
\end{align}
where we have explicitly listed the dependencies on dust size $a$. This is a Fredholm equation of the second kind with separable kernel, which can be solved explicitly:
\begin{align}
\Delta\ud(a) = \frac{\taus(a)}{\rho} \left(\nabla p - 2\rhog \eta \hatx\right),
\label{eqDeltaU}
\end{align}
with total density $\rho = \rhod + \rhog$. Multiplying by $\sigma/\rhod$ and integrating the equation for $\Delta\ud$ over size we obtain a size-integrated relative velocity
\begin{align}
\Delta{\bf v} = \vd-\vg = \frac{\bartaus}{\rho} \left(\nabla p - 2\rhog \eta \hatx\right),
\label{eqDeltaV}
\end{align}
with average stopping time
\begin{align}
\bartaus = \frac{1}{\rhod}\int \sigma \taus(a) \rmd a.
\end{align}

The evolution of the total momentum $\rho {\bf v} = \rhod\vd + \rhog\vg$ necessarily does not involve the drag force, while the stress tensor appearing in (\ref{eqDustMomInt}) is:
\begin{align}
\mathsf{S} &=  \int \sigma(\ud - \vd)(\ud + \vd) \rmd a\nonumber\\
&= \int \sigma(\Delta \ud- \Delta {\bf v})\left(2{\bf v} + \Delta \ud + \frac{\rhog - \rhod}{\rho}\Delta {\bf v}\right) \rmd a.
\end{align}
To lowest order in $\stokes$, the second factor in parenthesis is $2{\bf v}$, so that the TV approximation to the stress tensor is
\begin{align}
\mathsf{S} &= 2{\bf v}\left(\int \sigma\Delta \ud \rmd a - \rhod\Delta {\bf v}\right) = 0.
\end{align}
This, together with the fact that the drag force cannot appear in the total momentum equation, means that the total momentum equation is \emph{exactly} the same as in the monodisperse case, and reads, in the TV approximation, and therefore ignoring terms that are quadratic and higher in the relative velocity:
\begin{align}
\partial_t\left(\rho{\bf v}\right) +& \nabla\cdot\left(\rho{\bf v}{\bf v} \right) =\nonumber\\
& 2\eta \rhog \hatx - \nabla p -2\rho\bm{\Omega}\times {\bf v}-\rho\nabla\Phi.
\label{eqMomTot}
\end{align}
The evolution of the total density can be found by adding up (\ref{eqGasContApp}) and (\ref{eqDustContInt}):
\begin{align}
\partial_t\rho +& \nabla\cdot(\rho {\bf v}) = 0.
\label{eqContTot}
\end{align}
In addition, we need an equation for the pressure, which, in the case of an isothermal gas component, is set exclusively by the gas density and therefore the gas continuity equation (\ref{eqGasContApp}):
\begin{align}
\partial_t p +& \nabla\cdot(p \vg) = 0.
\end{align}
In terms of ${\bf v}$ and $\Delta {\bf v}$, we have that $\vg = {\bf v} - \rhod\Delta {\bf v}/\rho$. This gives rise to a 'cooling term' on the right hand side \citep{2017ApJ...849..129L}:
\begin{align}
\partial_t p +& \nabla\cdot(p {\bf v}) = \nabla\cdot \left(\frac{p\rhod\Delta {\bf v}}{\rho}\right).
\end{align}
Using our expression (\ref{eqDeltaV}) for $\Delta {\bf v}$, we obtain
\begin{align}
\partial_t p +& \nabla\cdot(p {\bf v}) = \nonumber\\
&\nabla\cdot \left(\left(1-\frac{p}{c^2\rho}\right)\frac{p\bartaus \left(\nabla p - 2\rhog \eta \hatx\right)}{\rho}\right).
\label{eqPres}
\end{align}
If we set $\eta=0$ and take the monodisperse limit so that $\bartaus=\taus(a_0)$ the right hand side is equivalent to equation (16) of \cite{2017ApJ...849..129L} if we recognize their definition of the relative stopping time $t_{\rm s}= \rhog\tau_s/\rho$.

For a polydisperse dust fluid, the cooling term depends on the size-averaged stopping time $\bartaus$, which depends on $\sigma$. We therefore need to include the dust continuity equation (\ref{eqDustContApp}), which reads, when inserting $\ud = \Delta \ud + \vg = \Delta\ud  + {\bf v} - \rhod\Delta {\bf v}/\rho$:
\begin{align}
\partial_t\sigma + \nabla\cdot(\sigma {\bf v})= -\nabla\cdot(\sigma \left(\Delta\ud - \rhod\Delta {\bf v}/\rho\right)).
\end{align}
With the expressions for the relative velocities (\ref{eqDeltaU}) and (\ref{eqDeltaV}) we find
\begin{align}
\partial_t\sigma +& \nabla\cdot(\sigma {\bf v})= \nonumber\\
&\nabla\cdot\left[\frac{\sigma}{\rho} \left(\nabla p - 2\rhog \eta \hatx\right)\left(\frac{\rhod}{\rho}\bartaus -\taus(a)  \right)\right].
\label{eqSigma}
\end{align}

The TV equations are then given by (\ref{eqContTot}), (\ref{eqMomTot}), (\ref{eqPres}) and (\ref{eqSigma}):
\begin{align}
\partial_t\rho + \nabla\cdot\left(\rho{\bf v}\right)&= 0\, ,\label{eq:totalDens}\\
\partial_t{\bf v} + ({\bf v}\cdot\nabla){\bf v}  &= \frac{2p\eta{\bf \hat x}}{c^2\rho} - \frac{\nabla p}{\rho} -2\bm{\Omega}\times {\bf v}-\nabla\Phi\, ,\\
\partial_tp + \nabla\cdot\left(p{\bf v}\right)&= \mathcal{C}_{\rm g}\, ,\\
\partial_t\sigma + \nabla\cdot\left(\sigma {\bf v}\right)&= \mathcal{C}_{\rm d}\, \label{eq:dustSize},
\end{align}
where the cooling terms \citep{2017ApJ...849..129L} are given by:
\begin{align}
\mathcal{C}_{\rm g} &= \nabla\cdot\left(\left(1-\frac{p}{c^2\rho}\right)\frac{p}{\rho}\bartaus\left(\nabla p - \frac{2p \eta{\bf \hat x}}{c^2}\right) \right),\\
\mathcal{C}_{\rm d} &= \nabla\cdot\left(\frac{\sigma}{\rho}\left(\fsd\bartaus-\taus(a)\right)\left(\nabla p - \frac{2p\eta{\bf \hat x}}{c^2}\right)\right).
\end{align}

\subsection{The TV PSI dispersion relation}
\label{app:tv_disperse}

The equilibrium background state in our unstratified shearing box consists of constant gas and dust (size-) densities, and an equilibrium centre-of-mass velocity is ${\bf v} = (-Sx-\fsg\eta/\Omega){\bf \hat y}$, where $\fsg=\rhog/\rho$ is the gas fraction. Consider small perturbations to equations (\ref{eq:totalDens})--(\ref{eq:dustSize}) such that $\rhog = \rhogn + \hatrhog\exp(\rmi {\bf k}\cdot {\bf x} - \rmi \omega t)$, where $\rhogn$ is the background state with $|\hatrhog| \ll \rhogn$, 
and similarly for other quantities, yielding:
\begin{align}
-\rmi \omega \hat\rho
+& \rmi {\bf k}\cdot {\bf \hat v} = 0,\\
-\rmi\omega{\bf \hat v}
-& S\hat v_x{\bf \hat y}
=
\gsube\hatx\left(\hat p  - \hat\rho\right)
- \rmi c^2\fsgn{\bf k}\hat p -2\bm{\Omega}\times {\bf \hat v},\\
-\rmi\omega\hat p  +& \rmi{\bf k}\cdot{\bf \hat v} =
-\bartaus^0 c^2k^2 \fsdn \fsgn\hat p \nonumber\\
-&\bartaus^0
\left(\rmi k_x \gsube\left[
 (2\fsdn-\fsgn) \hat p
+(\fsgn-\fsdn)\hat\rho
+\fsdn\hattaus\right]
\right),\\
\rmi\omega\hat\sigma -& \rmi{\bf k}\cdot{\bf \hat v} =
\left(f_d^0\bartaus^0-\taus(a)\right)\left(k^2c^2 \fsgn\hat p + \rmi k_x\gsube \left[\hat p +\hat\sigma - \hat\rho\right]\right) \nonumber\\
&+
\rmi k_x \gsube \bartaus^0\left(\fsdn \hattaus + \fsgn(\hat\rho - \hat p) \right),
\end{align}
with $\gsube = 2\fsgn \eta$ and perturbed stopping time
\begin{align}
\hattaus = \frac{\bartaus^1}{\bartaus^0} = \frac{1}{\bartaus^0\rhod^0}\int \hat\sigma \sigma^0(a)\taus(a) \rmd a -  \frac{\hat\rho - \fsgn\hat p}{\fsdn}.
\end{align}
A further simplification is possible if we consider the gas to be incompressible, which is a good approximation for monodisperse SI modes \citep{2007ApJ...662..613Y}. We arrive at the incompressible limit by neglecting all pressure perturbations unless they are multiplied by the sound speed \citep{2017ApJ...849..129L}:
\begin{align}
-\rmi \omega \hat\rho
+ \rmi {\bf k}\cdot {\bf \hat v} =& 0\, ,\label{eqPertRhoApp}\\
-\rmi\omega{\bf \hat v}
- S\hat v_x{\bf \hat y}
=&
-\gsube{\bf \hat x}\hat\rho
- \rmi c^2 \fsgn {\bf k}\hat p -2\bm{\Omega}\times {\bf \hat v}\, ,\label{eqPertVelApp}\\
\rmi{\bf k}\cdot{\bf \hat v} =&
-\bartausn
\left(c^2k^2 \fsdn \fsgn \hat p
+ \rmi k_x \gsube(\fsgn -\fsdn )\hat\rho\right)\nonumber\\
&-\rmi k_x \gsube \fsdn \hattaus\, ,\label{eqPertPresApp}\\
\rmi\omega\hat\sigma - \rmi{\bf k}\cdot{\bf \hat v} =&
\left(\fsdn\bartausn-\taus(a)\right)\left(k^2 c^2 \fsgn\hat p + \rmi k_x\gsube \left[\hat\sigma - \hat\rho\right]\right) \nonumber\\
&+
\rmi k_x \gsube \left(\fsd \hattaus + \fsgn \bartausn \hat\rho\right),\label{eqPertSigmaApp}
\end{align}
and
\begin{align}
\hattaus &= \frac{1}{\rhodn}\int \hat\sigma \sigma^0(a)\taus(a) {\rm d}a -  \frac{\bartausn\hat\rho }{f_d^0}\, .
\label{eqTauHatApp}
\end{align}
Note that in a monodisperse dust fluid $\hattaus=0$. 

Adding up equations (\ref{eqPertPresApp}) and (\ref{eqPertSigmaApp}) to eliminate $\hattaus$ we obtain an expression for $\hat\sigma$:
\begin{align}
\left(\wstar  + \frac{\taus(a)}{\bartausn}\right)\hat\sigma
&=
\frac{\taus(a)}{\bartausn}\left(\frac{\rmi k^2 c^2 \fsgn \hat p}{k_x\gsube} + \hat\rho\right)\, ,
\label{eq:sigma}
\end{align}
with $\wstar  = (\omega -\fsdn \bartausn k_x\gsube)/(k_x \gsube \bartausn)$. Use (\ref{eq:sigma}) in (\ref{eqTauHatApp}) to find
\begin{align}
\hattaus =
\bartausn \left(\frac{\rmi k^2c^2 \fsgn \hat p}{k_x\gsube} + \hat\rho\right) \mathcal{I}(\wstar )
- \frac{\bartausn \hat\rho}{\fsdn}\, ,
\end{align}
with integral
\begin{align}
\mathcal{I}(\wstar ) = \frac{1}{\rhodn}\int
\frac{\sigma^0(a)}{\wstar  + \frac{\taus(a)}{\bartausn}}\left(\frac{\taus(a)}{\bartausn}\right)^2 {\rm d}a.
\end{align}
Equations (\ref{eqPertRhoApp})-(\ref{eqPertSigmaApp}) can then be combined to yield a dispersion relation
\begin{align}
& \left(\frac{k^2}{k_z^2}\omega^2 -\kappa^2\right) (\omega  - \fsdn\bartausn k_x \gsube)
=  \nonumber\\
&\qquad  \bartausn \fsdn 
\left(\rmi \frac{k^2}{k_z^2} (\omega^2-\kappa^2)\omega^2
- k_x \gsube \kappa^2\right)  \left(1-\mathcal{I}(\wstar )\right).
\label{eq:TVdisp}
\end{align}

\begin{figure}
	\includegraphics[width=0.9\columnwidth]{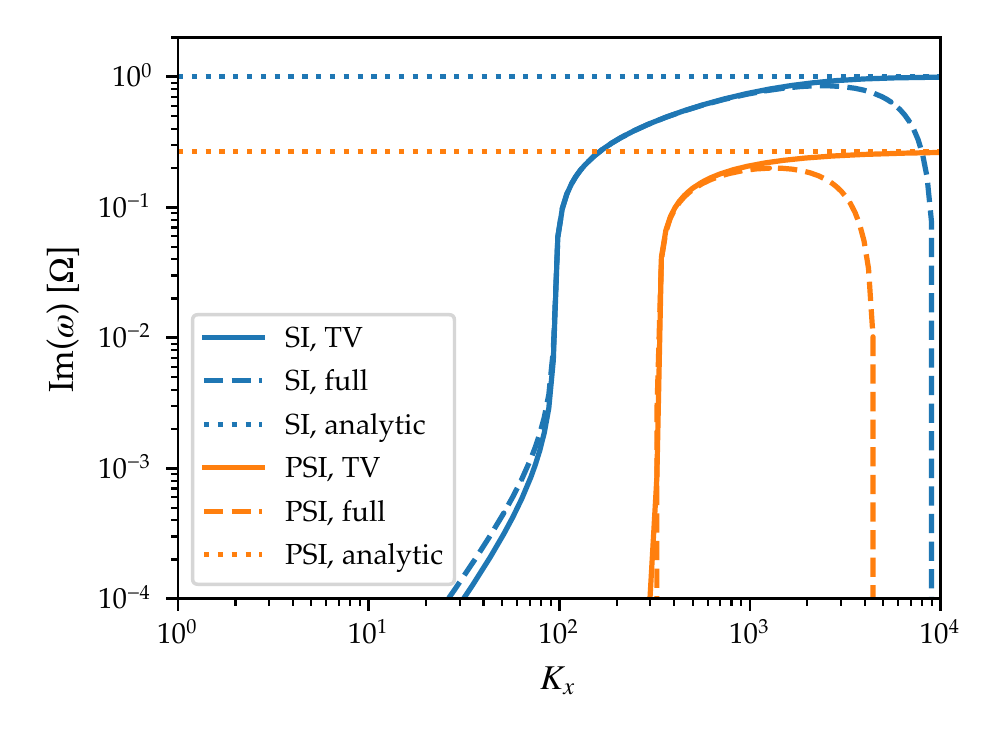}
    \caption{Variation of the growth rates with wave number of the monodisperse SI at stopping time $10^{-2}\Omega^{-1}$ (blue curves) and the PSI, using an MRN size distribution in stopping time range $[10^{-8}, 10^{-2}]\, \Omega^{-1}$ (orange curves), all for a dust to gas ratio $\mu=3$ with $K_z=K_x$. Shown are results for the full equations (dashed curves), the TV approximation (solid curves), and the analytical high wavenumber limit (dotted lines).}
    \label{fig:highmu}
\end{figure}

\subsection{Validity of the terminal velocity approximation}
\label{app:tv_valid}

The TV approximation holds if the perturbation and dynamical time scales are longer than the relative stopping time $t_{\rm s}$ and the length scales under consideration are longer than the stopping length $\eta t_{\rm s}^2$ \citep{2011MNRAS.415.3591J}. Formally, one requires the Stokes number $\stokes \ll 1$ \citep{2019MNRAS.488.5290L}, although good results have been reported on the monodisperse streaming instability up to $\stokes=0.1$ \citep{2017ApJ...849..129L}. In this work, we stick to a conservative limit of $\stokes < 0.01$ in order to remain well inside the TV regime. The wave number under consideration should satisfy $K <  \left(\fsgn \stokes\right)^{-2}$, so that the wave length of the perturbation is longer than the stopping length. While technically this makes the TV approximation a long wavelength approximation, in practice the main limitation usually comes from neglecting higher order terms in $\stokes$.

This is illustrated in Figure \ref{fig:highmu}, where we show the growth rate of the PSI as a function of wave number $K_x=K_z$ both for the full model and the TV approximation for the case with $\mu=3$ and stopping time range $[10^{-8}, 10^{-2}]\, \Omega^{-1}$. For comparison, we also show the growth rates of the monodisperse SI at stopping time $10^{-2}\, \Omega^{-1}$ for the same dust to gas ratio. We are clearly in the high-$\mu$ parameter range, where the unstable wave numbers have $K_x \stokes\gg 1$, while the low-$\mu$ SI is known to peak around $K_x \stokes \sim 1$ \citep{2005ApJ...620..459Y}. The TV approximation does a good job reproducing the full model up to $K_x \approx 10^3$, after which the full model shows a decline while the TV growth rate approaches the analytic estimate. The maximum $K_x$ so that the wavelength is longer than the stopping length is $\sim 10^5$, which means that based on this criterion the TV approximation should be valid across the whole domain shown in Figure \ref{fig:highmu}. It was shown in \cite{2018MNRAS.477.5011S} for the monodisperse SI that the failure of the TV model is due to the neglect of higher order terms in $\stokes$. Nevertheless, the analytic limit does a decent job of predicting the maximum growth rate of the full model to within $\sim 20\%$. The prediction gets better when the range of unstable wave numbers gets wider.

\section{Numerical Methods for Full Eigenproblem}
\label{app:numerical}

\begin{figure}
\includegraphics[width=\columnwidth]{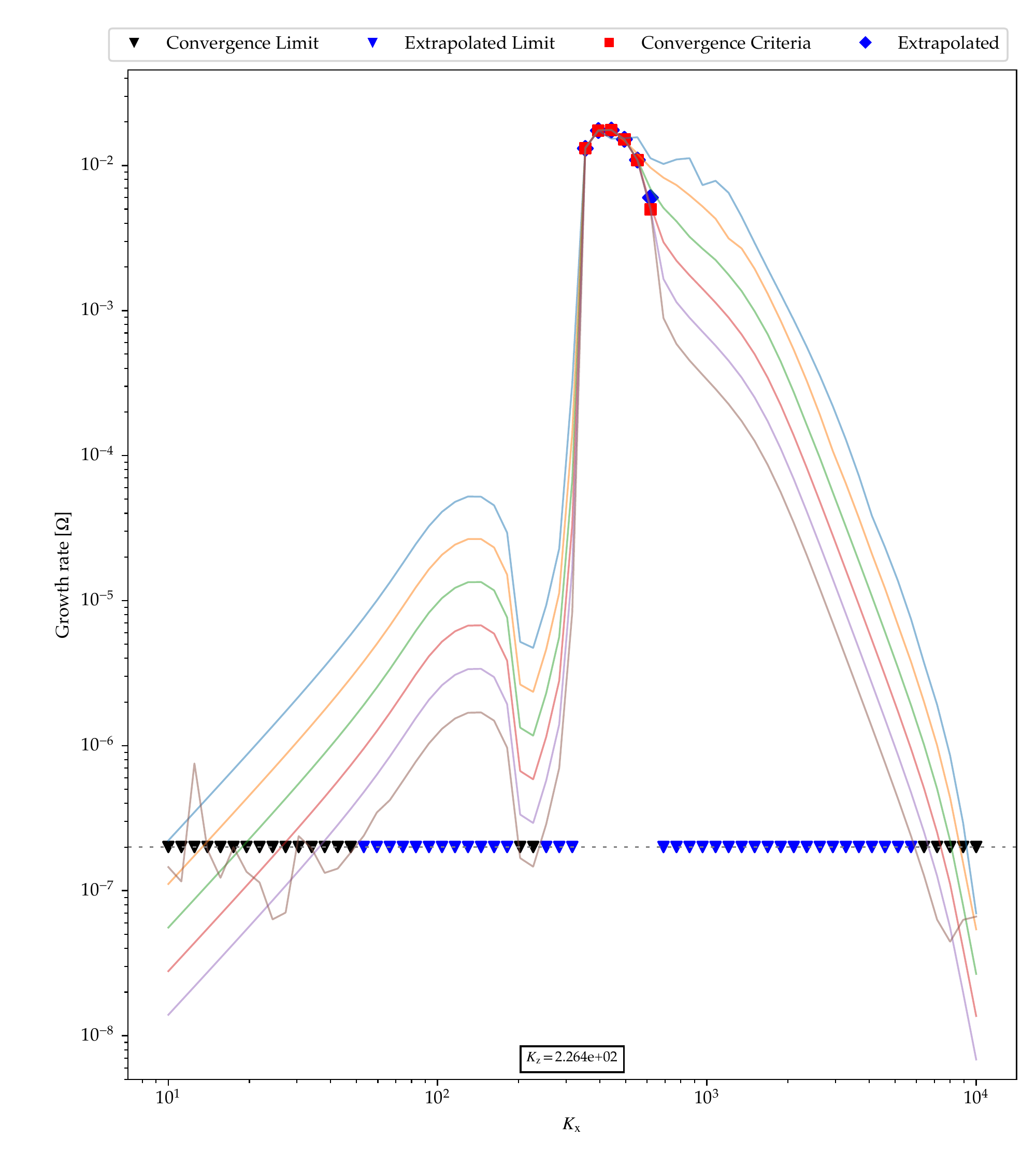}
\caption{Large scale single panel demonstrating the presentation used in Figures~\ref{fig:convmu0p5}--\ref{fig:convmu3} of detailed results of numerical calculations of the PSI growth rate. This instance for $\mu=0.5$, $\taus=[10^{-8},10^{-2}]$ at a range of vertical wavenumbers $K_z$.
{\sl Lines}: Numerical results at resolutions of $2^{[7,\dots,12]}+1$ points  in $\taus(a)$. 
{\sl Symbols}: Marks denoting the convergence decision for each $K_x$ parameter as per legend.}
\label{fig:convmu0p5zoom}
\end{figure}

\begin{figure*}
\includegraphics[width=\textwidth]{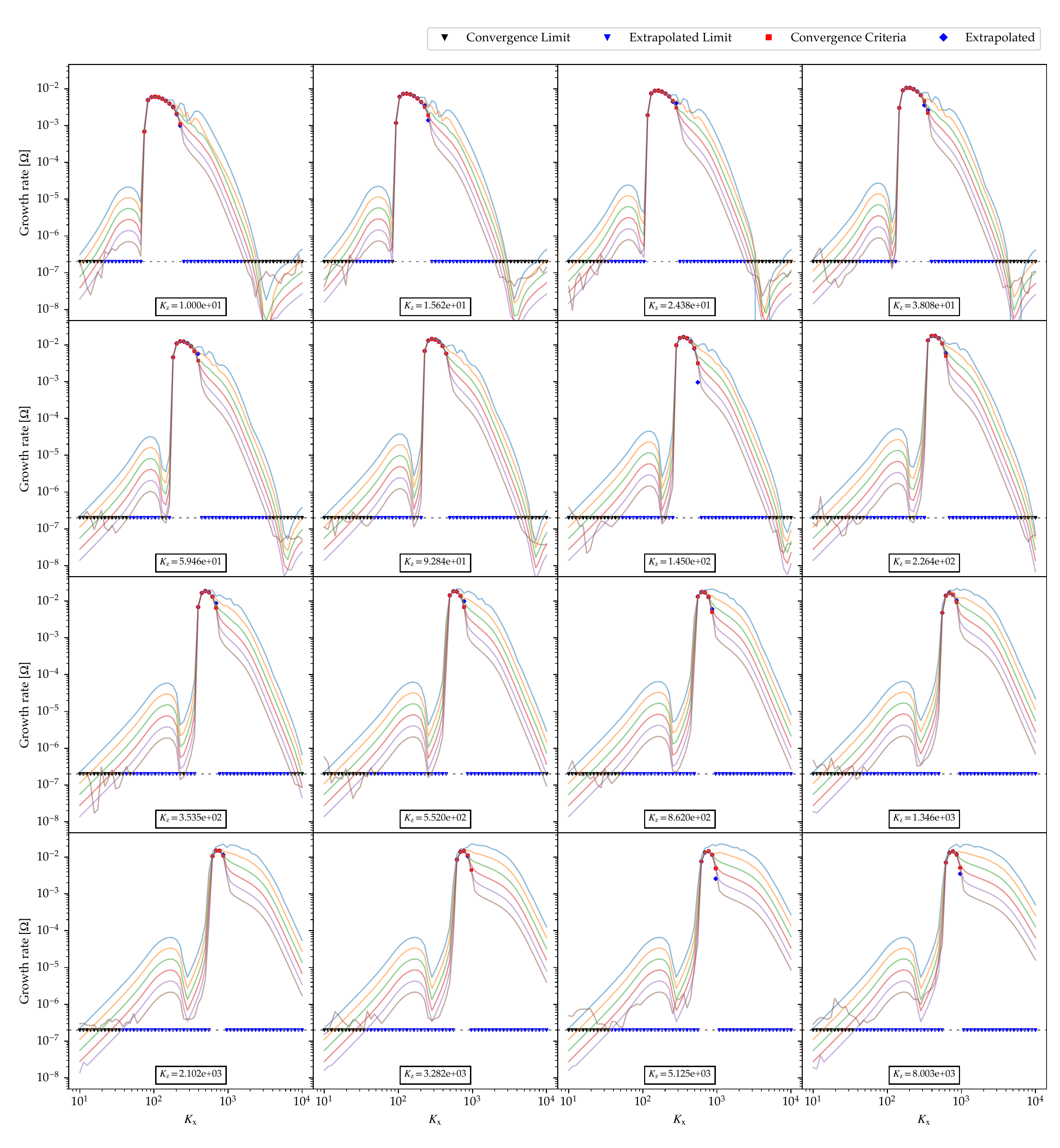}
\caption{Detailed results of numerical calculations of the PSI growth rate for $\mu=0.5$, $\taus=[10^{-8},10^{-2}]$ at a range of vertical wavenumbers $K_z$.
{\sl Lines}: Numerical results at resolutions of $2^{[7,\dots,12]}+1$ points  in $\taus(a)$. 
{\sl Symbols}: Marks denoting the convergence decision for each $K_x$ parameter as per legend.}
\label{fig:convmu0p5}
\end{figure*}

\begin{figure*}
\includegraphics[width=\textwidth]{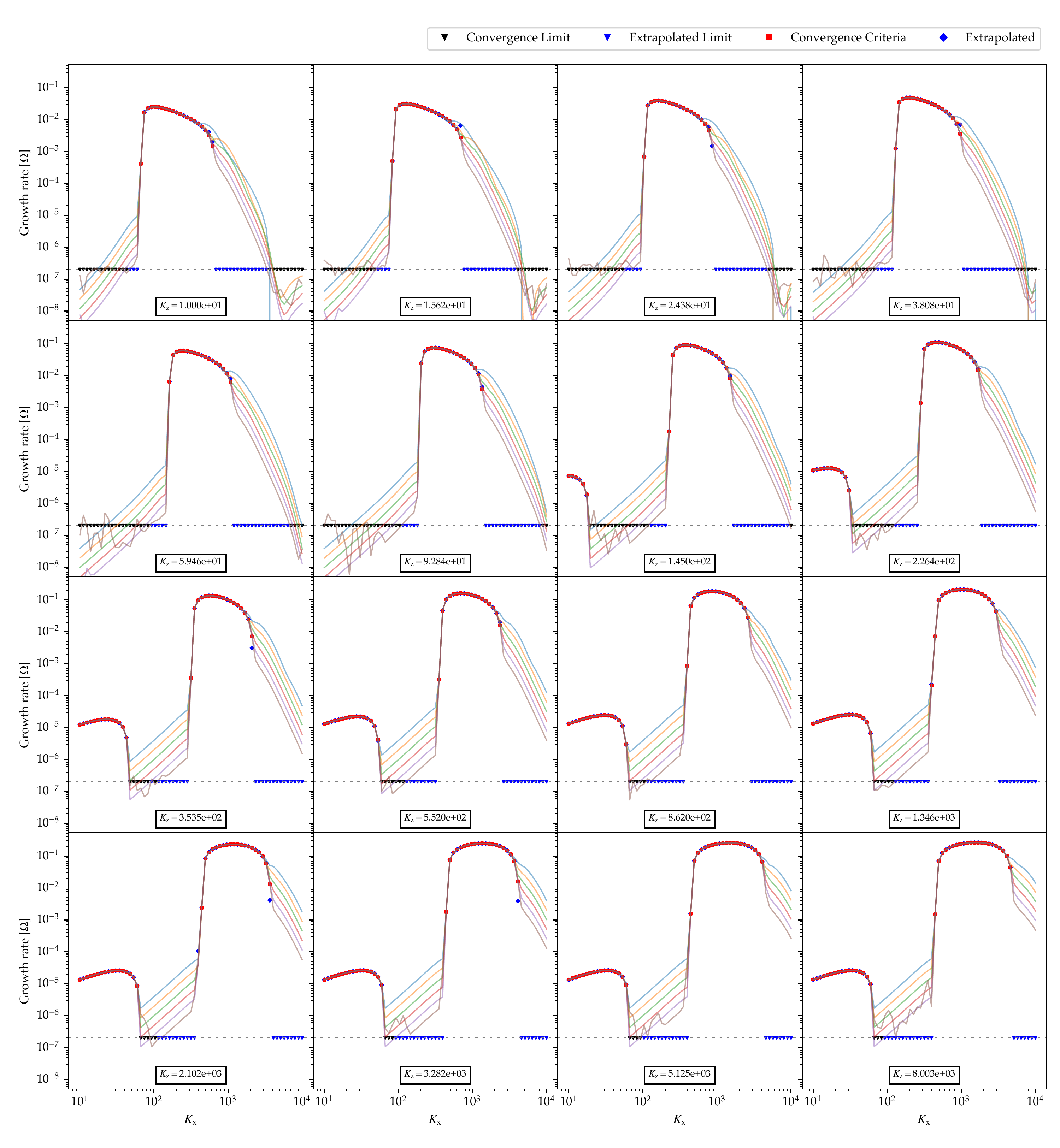}
\caption{Detailed results of numerical calculations of the PSI growth rate for $\mu=3$, $\taus=[10^{-8},10^{-2}]$.
{\sl Lines}: Numerical results at resolutions of $2^{[7,\dots,12]}+1$ points  in $\taus(a)$. 
{\sl Symbols}: Marks denoting the convergence descision for each $K_x$ parameter as per legend. Where both a symbol denoting that the two highest resolutions meet the convergence criteria and a mark for an extrapolated value appear, the value satisfying the convergence criteria is used.}
\label{fig:convmu3}
\end{figure*}

To solve the full eigenproblem for the PSI,
first, we choose the equilibrium gas and dust densities $\rhogn$, $\sigma^0(\taus)$.
Then, Equations (17)--(20) can be solved for the background state gas velocity $\vgxn$, $\vgyn$, $\vgzn$, 
and the dust velocity $\uxn(\taus)$, $\uyn(\taus)$, $\uzn(\taus)$ by numerical quadrature.
To solve the  eigenvalue problem (21)--(24) for eigenvalues $\omega$  we
discretize the dust eigenfunctions  $\hatsigma(\taus)$ and ${\bf \hat{u}}(\taus)$  by sampling at points on a 
Chebyshev grid \citep[][eq.~A.19]{Boyd} in the interval [$\tau_{\rm s,min}$,\ $\tau_{\rm s,max}$] 
on $L$ points.
This transforms Equations~(23)--(24) into $4L$ scalar equations.
In the  monodisperse case, the  Fourier analyzed  compressible SI problem
produces a matrix eigenvalue problem in six scalar variables which can be solved for six eigenvalues and eigenvectors of length 6.
In contrast, this discretization of the PSI eigenproblem produces $4+4L$ eigenvalues $\omega_{L,0\dots 4+4L}$ and eigenvectors of the corresponding length.
Many of these are numerically spurious, meaning that they do not correspond to an eigenvalue of the continuous problem,
 and have decay rates which grow with $L$ \citep{Boyd}.
The eigenvectors in turn contain the Fourier coefficients of the gas density and velocity eigenfunctions, 
and discretized coefficients as a function of $\taus$ 
of the dust density eigenfunctions $\hatsigma_{L,i}(\taus)$ and dust velocity eigenfunctions ${\bf \hat{u}}_{L,i}(\taus)$.
As $L$ is increased, these successive approximations to the physical eigenfunctions, of which we are concerned with the fastest growing, converge towards the exact result, 
\begin{align}
\lim_{L\rightarrow \infty} \hatsigma_L(\taus) = \hatsigma(\taus)\, .
\end{align}
We discretize all the integral terms in Equation 22 over $\taus(a)$ with a trapezoid rule quadrature which allows the entire discretized eigenproblem to be expressed in a single $4+4L$ by $4+4L$ matrix.
The convergence properties of trapezoid rule quadrature means that when the dust eigenfunctions $\hatsigma(\taus)$ and ${\bf \hat{u}}(\taus)$ are 
smooth functions of $\taus$, the asymptotic convergence rate of the approximation for the eigenvalues $\omega$ will be second order.
However, when any of these eigenfunctions are non-smooth, the method will converge at only first order.
Thus, when convergence is slow, we exploit a series of results to obtain a more precise one.

In essence, Richardson extrapolation consists of fitting a polynomial to the partial sums of a series 
to produced a transformed series with accelerated convergence properties.
 It underlies some common 
numerical procedures, such as Romberg integration, where it is used to accelerate the convergence 
of trapezoid rule quadrature, and as a general tool for uncertainty quantification 
in the verification of numerical simulations with PDEs \citep{roache1998verification}.
We employ the N-step algorithm for Richardson extrapolation described by \citet[][p. 375]{1978amms.book.....B}
applying it to the imaginary components of the fastest growing eigenvalues, the set $g_L= \{\max_i(\mathcal{I}(\omega_{L,i}))\}$ for $L\in2^{[7,\dots,12]}+1$.
Applied to the latter half of the series the polynomial is expression for the extrapolated value $g_{\rm R}$ is
$g_{\rm R} = 2 g_{513} -9 g_{1025} + 8 g_{2049}$,
 involving a zero weight on the highest resolution result $g_{4097}$, facilitating comparison between the extrapolated and highest resolution result as another indication of the residual error.
 The eigenvalue computations are performed with {\tt numpy.linalg.eig} from the Intel Python Distribution.

Finally, we report only growing modes with a well defined sense of convergence, and only upper limits on growth for 
other cases. Detailed sections of the parameter grids, the raw computations, and annotations about the 
convergence criteria are presented as an larger scale example key to the symbols in Figure~\ref{fig:convmu0p5zoom}
and for more cross sections in Figures~\ref{fig:convmu0p5}--\ref{fig:convmu3}.
In general, the physically most relevant growing modes converge quickly, while convergence in 
cases where no significant growth is found is much slower, but regular.
 From inspection of the results in  Figures~\ref{fig:convmu0p5}~and~\ref{fig:convmu3}, a typical value below which floating point accuracy corrupts results is a growth rate of $2\times10^{-7}$.
We conservatively choose to present a reasonably well converged directly computed value over those results 
employing Richardson extrapolation to accelerate convergence.
This convergence criteria is a relative error between the two highest resolution computations of $5\%$ in imaginary part, and $10\%$ in real part.

For the series of computations with varying $\taus$ resolutions $L\in2^{[7,\dots,12]}+1$ at a fixed wavevector $\bf k$ the sequence of criteria used to determine the result shown is:
\begin{enumerate}
\item Are the growth rates at any resolution  $<\ 2 \times 10^{-7}$? Decision: Accept upper limit $2 \times 10^{-7}$.
\item Are the two highest resolution results within the error tolerance? Decision: Accept highest resolution result.
\item Is the Richardson extrapolation of the converging series of results $\geq 2 \times 10^{-7}$? Descision: Accept Richardson extrapolated value.
\item Is the Richardson extrapolation $<2\  \times 10^{-7}$? Descision: Accept upper limit $2 \times 10^{-7}$.
\end{enumerate}
Each of these criteria is evaluated in order until a decision is accepted.

\section{Comparison to Krapp~et~al.~(2019)}
\label{app:krapp}

\begin{figure}
\begin{center}
\includegraphics[width=0.8\columnwidth]{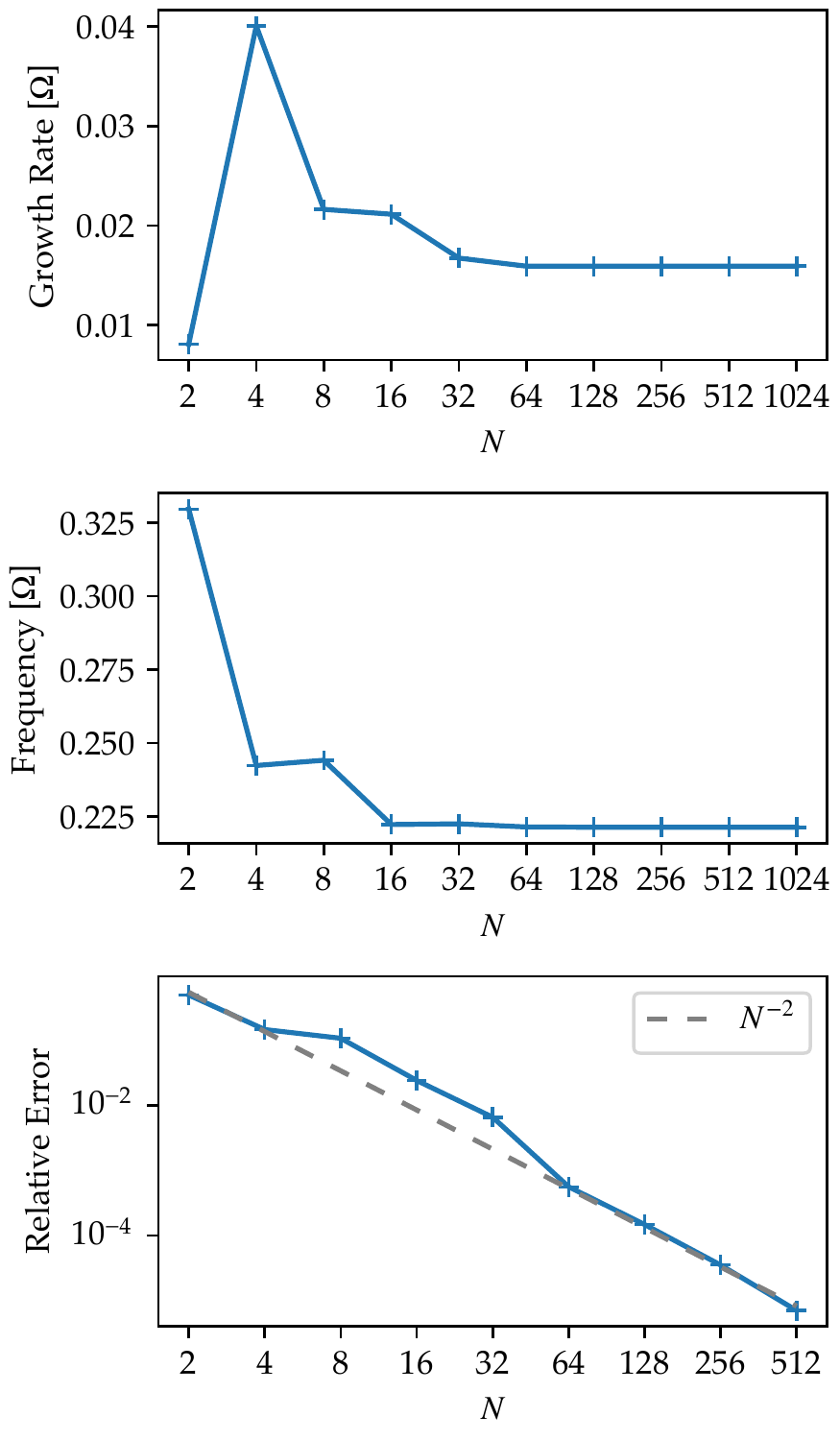}
\end{center}
\caption{Convergence study on the mode we label ${\bf K}_{\rm fast,128}$. Top panel: growth rate (imaginary part of $\omega$ in our notation), middle panel: oscillation frequency (real part of $\omega$ in our notation), bottom panel: relative error compared to the $N=1024$ result, showing the second order convergence of the direct solver to an eigenvalue with growth.}
\label{fig:krappfig2}
\end{figure}

\begin{table*}
	\centering
	\caption{Numerical results for the maximum growth rate in the range $0.1 \leq K_x,K_z \leq 1000$ for an MRN size distribution. Columns from left to right: dust to gas ratio, stopping time range, wavenumber $x$, wavenumber $z$, mode frequency as found from the eigensolver, and the mode frequency as found from the terminal velocity solver.}
	\label{tab:krapp2}
	\begin{tabular}{cccccc} 
		\hline
		$\mu$ & $\Omega \taus$ & $K_x$ & $K_z$ & $\nu$ & $\nu_{\rm TV}$\\
		\hline
		$1$ & $\left[10^{-4},10^{-3}\right]$ & $1000$ & $195.13$ & $0.1208+0.04277\rmi$ & $0.1208+0.04281\rmi$\\
		$1$ & $\left[10^{-4},10^{-2}\right]$ & $666.91$ & $1000$ & $0.6181+0.1190\rmi$ & $0.6194+0.1414\rmi$\\
		$0.5$ & $\left[10^{-4},10^{-3}\right]$ & $1000$ & $96.154$ & $0.06565+0.01331\rmi$ & $0.06566 + 0.01338\rmi$\\
		$0.5$ & $\left[10^{-4},10^{-2}\right]$ & $631.52$ & $981.70$ & $0.6809+0.03178\rmi$ & $0.6819+0.06040\rmi$\\		
		\hline
	\end{tabular}
\end{table*}

A similar problem to the one studied in this paper was presented in \citetalias{2019ApJ...878L..30K}. We have outlined the differences in the numerical approaches in section \ref{sec:numerical}. To reiterate: both \citetalias{2019ApJ...878L..30K} and the present work aim at studying a continuous size distribution, but the underlying physical model is different at finite resolution. At finite resolution, \citetalias{2019ApJ...878L..30K} solve for the exact growth rates for a system consisting of a finite number of single-sized dust fluids, while in this work we calculate the approximate growth rates for a continuous size distribution. Both methods should give the same answer in the limit of infinite resolution in size space (modulo the different drag law used in \citetalias{2019ApJ...878L..30K}, the effect of which should be small). However, at finite resolution, differences are to be expected, for example due to the different equilibrium states (see section \ref{sec:numerical}). Meaningful comparisons of growth rates can therefore only be made in the continuum limit. We note that since both methods solve for the eigenvalues of a dense matrix, the computational effort should be similar for the same $N$ (number of species/collocation points).

In this Appendix, we provide additional comparisons and convergence results. We do note first of all that our results agree qualitatively (compare the top right panel of Figure 1 of \citetalias{2019ApJ...878L..30K} with the middle right panels of Figure \ref{fig:kxkz}), but a more detailed comparison over all wave numbers is not meaningful because differences between the middle and right columns of their Figure 1 show that parts of $K$-space have not converged.

The one point in $K$-space depicted in Figure 1 of \citetalias{2019ApJ...878L..30K} where we do know for certain that they obtain a result that converges in the continuum limit, and therefore lends itself to detailed comparison, is at the location of the white triangle in the top middle panel of their Figure 1. Let us call the wave number of the triangle ${\bf K}_{\rm fast,128}$. The convergence history of the fastest growing mode is depicted in their Figure 2, and it appears that at a number of species $N_{\rm spec} = 128$ the mode frequency is converged enough so that the wave number of the fastest growing mode is likely to be ${\bf K}_{\rm fast,128}$ for $N_{\rm spec} \geq 128$. Note that this is probably not the case for $N_{\rm spec} < 64$, and definitely not the case for  $N_{\rm spec} = 16$, as apparent from ${\bf K}_{\rm fast,16}$, indicated by the position of the white triangle in the top left panel of their Figure 1. It seems that the mode at ${\bf K}_{\rm fast,16}$ does not reach a converged value for $N_{\rm spec} = 512$, which means no meaningful comparison can be done. We therefore have to limit our comparison to ${\bf K}_{\rm fast,128}$. We have extracted ${\bf K}_{\rm fast,128}$ from a digitized version of the top middle panel of Figure 1 of \citetalias{2019ApJ...878L..30K}, and obtained the mode frequency from a digitized version of the left two panels of their Figure 2, obtaining ${\bf K}_{\rm fast,128} = (20.6819, 6.40829)$. We compare our results with theirs in Table \ref{tab:krapp}, showing agreement to $3\%$. We show the convergence history for the growth rate and oscillation frequency in the top two panels of Figure \ref{fig:krappfig2}. It should be stressed that these are at fixed wave number, while it is likely that the wave numbers in Figure 2 of \citetalias{2019ApJ...878L..30K} vary with the number of species. In the bottom panel of Figure \ref{fig:krappfig2} we show that the relative error decreases with of the number of collocation points as $N^{-2}$, as would be expected for the trapeziod rule quadrature. 

We also report that, while outside the TV regime which is the scope of this paper, our eigensolver does not yield a growing mode at the position of the white triangle in the bottom middle panel of Figure 1 of \citetalias{2019ApJ...878L..30K}, which is consistent with their results.  

In addition, in Table \ref{tab:krapp2} we provide additional benchmarks for the fastest growing modes for 4 of the columns shown in Figure 4 of \citetalias{2019ApJ...878L..30K} (left two columns of the top right and top right middle panels of the top row of their Figure 4). These results were obtained by using the dual annealing method \citep{1997PhLA..233..216X} as implemented in {\tt scipy.optimize.minimize} on our eigensolver to find the maximum growth rate in the domain $0.1 \leq K_x,K_z \leq 1000$, the same domain as considered in \citetalias{2019ApJ...878L..30K}. The values quoted are the digits which do not change between $N=512$ and $N=1024$, and the growth rates appear to be consistent with the colors shown in Figure 4 of \citetalias{2019ApJ...878L..30K}. We also quote the results of the TV solver, which gives good agreement except in the bottom row of Table \ref{tab:krapp2}, where it is off by a factor of $\sim 2$. A similar discrepancy can be observed in the top row, middle two panels of Figure \ref{fig:kxkz} towards the maximum $K_z$, and is probably due to the neglect of higher order terms in $\stokes$ in the TV approximation \citep{2018MNRAS.477.5011S}.

We therefore conclude that our results are consistent with those of \citetalias{2019ApJ...878L..30K} in all cases we have considered.  

\end{document}